\documentclass[twocolumn,10pt]{article}
\usepackage[top=2cm, bottom=2cm, left=1.75cm, right=1.75cm]{geometry}
\usepackage{times}

\usepackage{bbold}
\usepackage{amsmath}
\usepackage{flushend}
\usepackage{algorithm}
\usepackage{algpseudocode}
\usepackage[dvipsnames]{xcolor}
\usepackage[hyphens]{url}
\usepackage{paralist}
\usepackage{graphicx}		%
	\setkeys{Gin}{width=\textwidth, height=\textheight, keepaspectratio}

\usepackage{booktabs}
\usepackage[square,sort&compress,comma,numbers]{natbib}
\usepackage[labelfont=bf]{caption}
\usepackage{amssymb}
\usepackage{pifont}
\usepackage{multirow}
\usepackage{tablefootnote}

\usepackage[bookmarks=true, bookmarksnumbered=true, colorlinks=true, linkcolor=linkcol, citecolor=citecol, urlcolor=urlcol, hypertexnames=true]{hyperref}

\definecolor{darkgreen}{RGB}{0, 100, 0}
\definecolor{linkcol}{rgb}{0.3,0,0}
\definecolor{citecol}{rgb}{0.3,0,0}
\definecolor{urlcol}{rgb}{0.3,0,0}
\definecolor{vlightgray}{gray}{0.925}

\let\OLDthebibliography\thebibliography
\renewcommand\thebibliography[1]{
  \OLDthebibliography{#1}
  \setlength{\parskip}{0pt}
  \setlength{\itemsep}{1pt plus 0.2ex}
}

\usepackage[hang,flushmargin]{footmisc}

\makeatletter
\def\url@leostyle{%
  \@ifundefined{selectfont}{\def\UrlFont{}}%
  {\def\UrlFont{}}%
}
\makeatother
\usepackage[hyphenbreaks]{breakurl}

\newcommand{\descr}[1]{\smallskip\noindent\textbf{#1}}

\usepackage[raggedright,compact]{titlesec}
\titlespacing*{\section}{0pt}{*3.5}{3.5pt}
\titlespacing*{\subsection}{0pt}{*2.5}{2.5pt}
\titlespacing*{\subsubsection}{0pt}{*2}{2pt}

\usepackage[skip=0pt,font=scriptsize]{subcaption}

\usepackage{titling}
\newif\ifrevision
\revisiontrue
\revisionfalse
\ifrevision
	\newcommand{\revision}[2]{{\color{red} #2}}
\else
	\newcommand{\revision}[2]{#2}
\fi

\begin{document}
\title{\bf Revealing The Secret Power: How Algorithms Can Influence Content Visibility on Twitter/X\thanks{Published in the Proceedings of the 33rd Network and Distributed System Security Symposium (NDSS 2026) -- please cite the NDSS version.}}

\author{Alessandro Galeazzi$^1$, Pujan Paudel$^2$, Mauro Conti$^{1,3}$, Emiliano De Cristofaro$^4$, Gianluca Stringhini$^2$\\[1ex]
$^1$University of Padua, $^2$Boston University, $^3$Örebro University, $^4$UC Riverside}

\date{}

\maketitle

\begin{abstract}
In recent years, the opaque design and the limited public understanding of social networks' recommendation algorithms have raised concerns about potential manipulation of information exposure. 
Reducing content visibility, aka \emph{shadow banning}, may help limit harmful content; however, it can also be used to suppress dissenting voices. 
This prompts the need for greater transparency and a better understanding of this practice.

In this paper, we investigate the presence of visibility alterations through a large-scale quantitative analysis of two Twitter/X datasets comprising over 40 million tweets from more than 9 million users, focused on discussions surrounding the Ukraine–Russia conflict and the 2024 US Presidential Elections.
We use view counts to detect patterns of reduced or inflated visibility and examine how these correlate with user opinions, social roles, and narrative framings. 
Our analysis shows that the algorithm systematically penalizes tweets containing links to external resources, reducing their visibility by up to a factor of eight, regardless of the ideological stance or source reliability.
Rather, content visibility may be penalized or favored depending on the specific accounts producing it, as observed when comparing tweets from the Kyiv Independent and RT.com or tweets by Donald Trump and Kamala Harris.
Overall, our work highlights the importance of transparency in content moderation and recommendation systems to protect the integrity of public discourse and ensure equitable access to online platforms.
\end{abstract}

\section{Introduction}\label{sec:intro}
Digital spaces play a critical role in shaping today's information environment. 
Across the world, users rely on social platforms to access information on a wide range of topics, from personal interests to issues of societal importance.
Within this ecosystem, recommendation algorithms are pivotal in governing the visibility and circulation of content. %
These algorithms suggest or outright select content for users based on interests inferred from, e.g., their interactions and connections, as well as the content they create or engage with.

Crucially, recommendation algorithms can also limit the visibility of some content, e.g., hashtags related to dangerous or inappropriate topics or information that may encourage harmful behaviors~\cite{savolainen2022shadow}. 
This is known as ``{\em shadow banning},'' ``visibility alteration,'' or ``reduction,'' and typically aims to moderate online discussions, promote the dissemination of reliable content, and minimize the likelihood of toxic interactions to foster a healthier information ecosystem~\cite{jiang2023trade}.
\revision{C2c}{In practice, shadow banning and, in general, algorithmic alteration aimed at reducing the visibility of a user/their content might occur to control platform dynamics and for a number of reasons, ranging from mitigating spam/bot activity, algorithmic cleaning (e.g., throttling low-performing, repetitive, or self-promotional content), to prioritizing user retention (e.g., by penalizing posts linking to competing platforms).}

Alas, the details of how shadow banning works and the rationale behind penalizing procedures remain mostly opaque. 
In some cases, altering visibility may be a deliberate action by a platform's moderation team~\cite{yt_improving, nyt2025muskloomer}; in others, it could result from certain types of content being automatically flagged~\cite{gillespie2022not}. 
Whether humans or automated systems make these decisions, the lack of transparency surrounding shadow banning may introduce a certain degree of arbitrariness, real or perceived, as users often have little or no way of determining whether their content has been subject to visibility restrictions~\cite{jhaver2023personalizing}.

This opacity %
also encourages doubts as to whether or not shadow banning may be used to suppress specific viewpoints or communities~\cite{nyt2025muskloomer}. 
The difficulty in detecting it, compared to more visible actions like account suspensions or bans, further exacerbates these concerns. 
Users may remain unaware that their content is being deliberately suppressed, which can lead to accusations of bias or censorship~\cite{le2021setting}.
For example, journalists have reported internal mechanisms on platforms like Twitter (now X) and Facebook that reduce the visibility of certain types of content without notifying users~\cite{wp_shadow}.

While Twitter/X previously denied engaging in shadow banning~\cite{le2021setting}, the platform's updates to its terms of service in 2024 began to acknowledge the use of such practices~\cite{twitter_tos}. 
This has fueled debates about the fairness and transparency of content moderation.
The ability to shape or manipulate online debates to serve corporate or political interests arguably poses a significant threat to democratic discourse.
Moreover, some have criticized the potential for platforms to silence political or ideological views under the guise of content moderation~\cite{dunn2024online} -- for instance, Elon Musk claimed that Twitter previously penalized conservative viewpoints, asserting that the so-called ``Twitter Files'' suggested algorithmic interventions biased against certain ideologies~\cite{nytimes_musk}.

Beyond anecdotal evidence, researchers have begun looking into shadow banning through technical and statistical analyses (see Section~\ref{sec:related}). 
Overall, prior work suggests that shadow banning is relatively widespread across social media platforms~\cite{le2021setting,bartley2021auditing}.
However, the lack of transparency in applying these techniques raises serious concerns.
This prompts the need for systematic techniques to detect and measure visibility alterations and assess whether platforms employ shadow banning to favor specific narratives, limit dissenting voices, \revision{}{or prioritize corporate interests.}

\descr{Technical Roadmap.} \revision{C4}{Prior work has investigated shadow banning by analyzing search results or auditing recommendation algorithms, revealing potentially systematic penalization of specific users or content. 
However, these methods have not been able to determine whether content is penalized based on specific features (e.g., the presence of URLs), particularly at the debate level or across different topics.}

In this paper, we aim to address this gap and evaluate the presence of visibility alterations in the wild using {\em view counts,} a largely underexplored metric in related work introduced on Twitter in 2022 and once available through the now mostly defunct API.
As academic access to the API has been restricted since June 2023%
~\cite{rathje2024tackle,jaursch2024enabling}, we turn to two publicly available datasets that include view counts: 1) almost 35M tweets related to the 2024 US Presidential Election, released by Balasubramanian et al.~\cite{balasubramanian2024public}, and 2) over 17M tweets related to military aid in the Ukraine-Russia war, released by Baqir et al.~\cite{baqir2025unveiling}.\footnote{As Twitter rebranded to X in July 2023, tweets are now called posts. Since one of our datasets was collected before the rebranding and one after, to ease presentation, in the rest of the paper, we stick to ``tweets.''}
Our methodology relies on data analysis and network science techniques to uncover traces of shadow banning at the content, user, and network levels, \revision{C1}{shedding light on visibility alterations. 
Although the inaccessibility of the platform’s recommendation algorithm prevents us from establishing an explicit causal link to algorithmic design, our findings provide reasonable evidence in support of this hypothesis. 
While alternative explanations are possible, we discuss why they are unlikely to hold in Section~\ref{sec:discussion}.}

\descr{Research Questions.} Overall, we identify and address the following research questions:
\begin{itemize}
\item \textit{RQ1.} {\em Does the visibility of content depend on the characteristics of the information it contains?}
Our analysis examines how various content features, including the presence of external links, political bias, factuality of the sources, and ideological stance, may influence visibility.
\item \textit{RQ2.} {\em Do users experience different levels of visibility?}
We analyze the activities of prominent accounts to detect systematic variations in individual content visibility, i.e., whether some users' content consistently receives less exposure.
\item \textit{RQ3.} {\em Does visibility vary across different communities?}
Social media users tend to form ideologically homogeneous communities through repeated interactions; thus, we investigate whether the content generated within these communities exhibits varying levels of visibility.
\end{itemize}

\descr{Main Findings.} 
We introduce and use a new metric that accounts for author popularity when evaluating content visibility, called \emph{p-score}, testing its dependence on content, users, and community characteristics.
We find that visibility can vary substantially depending on content characteristics, both for content related to the Ukraine-Russia War and the 2024 US Elections.
Even users with comparable profiles experience drastic differences in visibility---sometimes differing by orders of magnitude.

Among other things, we emphasize the following findings:
\begin{enumerate}
\item At the content level (RQ1), tweets containing links to external websites have a significantly lower median p-score than those without links---respectively, approximately eight times for Ukraine-Russia and four times for the US Elections dataset.
\item From the point of view of user-level visibility (RQ2), accounts that frequently post URLs and automated accounts are further penalized.
Furthermore, in the Ukraine-Russia dataset, two prominent news outlets with comparable popularity (namely, RT.com and The Kyiv Independent) exhibit a median p-score differing by one order of magnitude (0.0069 vs. 0.084). 
A similar trend emerges in the US Elections dataset, where Donald Trump's median p-score is approximately four times higher than that of Kamala Harris (0.132 vs. 0.030).
\item Analyzing visibility variations at the community level (RQ3), using retweet and reply information, does not reveal systematic penalization of specific groups. %
\end{enumerate}

These discrepancies support the hypothesis that algorithmic interventions like shadow banning may be occurring, as Twitter/X's recommendation algorithm systematically favors tweets without external URLs.
Although we do not find evidence that specific topics, viewpoints, or communities are systematically penalized, interesting individual-level discrepancies do emerge, e.g., in the cases of The Kyiv Independent vs.~RT.com and Donald Trump vs.~Kamala Harris.

Overall, our work introduces a data-driven methodology using view counts and data science tools for detecting alterations in content exposure across different contexts.
It also further highlights the importance of accessing information about content exposure to enhance transparency in social media algorithms and promote fairness and trustworthiness in online information ecosystems.

\revision{C2d}{{\em Disclaimer:} Throughout the paper, we refer to systematic suppression, penalization, etc., in a broad sense, also to encompass selective content treatment.
That is, we do not imply that suppression is necessarily ideologically motivated.}

\section{Related Work}\label{sec:related}
In recent years, researchers have begun to work on identifying and categorizing various types of algorithmic moderation techniques, examining their consequences on the dynamics of information environments, and exploring their broader societal impact. 
Several studies highlight how algorithmic moderation shapes the visibility of content and influences public discourse. %
For instance, Zannettou et al.~\cite{zannettou2019let} present a large-scale analysis of Twitter, identifying content suppression instances, particularly in political debates, where specific hashtags were systematically downranked, limiting their reach. 
Jiang et al.~\cite{jiang2020reasoning} explore the impact of political bias in comment moderation on YouTube, finding that the likelihood of comment removal is independent of the political leanings of the video, even though other potential biases may persist in moderation practices.

Prior work has also examined the effects of algorithmic curation on users' exposure to diverse content. 
Algorithms have been shown to influence what users see on Twitter~\cite{bartley2021auditing}, Facebook~\cite{bakshy2015exposure}, TikTok~\cite{vombatkere2024tiktok}, YouTube~\cite{hosseinmardi2024causally}, and Google~\cite{robertson2018auditing}. 
Bakshy et al.~\cite{bakshy2015exposure} analyze how Facebook's algorithm reduces exposure to ideologically diverse news, contributing to filter bubbles and echo chambers~\cite{cinelli2021echo}. 
On TikTok, Vombatkere et al.~\cite{vombatkere2024tiktok} find that the recommendation system primarily favors content based on likes and friendship relations, potentially contributing to users' segregation.

In an effort to improve transparency, some platforms may offer explanations for why they remove content. 
However, these are often implausible or misleading, as observed by Mousavi et al.~\cite{mousavi2024auditing} for TikTok. 
Whereas, Jhaver et al.~\cite{jhaver2019does} demonstrate that providing users with meaningful explanations for content moderation decisions on Reddit reduces the likelihood of future post removals, indicating that transparency can positively influence user behavior.

Specific to shadow banning, Le Merrer et al.~\cite{le2021setting} investigate Twitter algorithmic interventions to reduce the visibility of certain content or accounts without outright removal.
They highlight widespread traces of visibility interventions across various user groups and communities.
Compared to our work, their study focuses on account-level interventions based on appearances in suggestions, search results, or discussion cascades, without considering alterations at the content level.

Beyond technical analysis, scholars have also examined the broader implications of algorithmic moderation, particularly its impact on free speech, public discourse, and user trust.
Gillespie~\cite{gillespie2022not} argues that content recommendation itself functions as a form of moderation, emphasizing the lack of transparency in how platforms manage visibility. 
This opacity can obscure biases in content curation and contribute to inequities among different communities, eroding trust in these systems.
Others have explored societal and political consequences. 
For example,  DeVito~\cite{devito2017editors} analyzes the limited information released about Facebook's News Feed algorithm, uncovering the underlying values that shape its decisions and calling for greater regulatory oversight. 

Overall, previous studies analyzing the presence of shadow banning and biases in algorithmic suggestion systems have mostly relied on bots~\cite{hosseinmardi2024causally}, developed auditing models~\cite{robertson2018auditing, mousavi2024auditing}, or examined algorithmic interventions by analyzing search query results~\cite{le2021setting}. 
However, algorithmic interventions that shape content visibility exist at different levels and in various forms, ranging from downranking in suggestion systems to complete hiding in search results. 
In this paper, we approach this issue from a broader perspective, leveraging the number of times a piece of content has been viewed to detect the presence of shadow banning. 
This allows us to capture the effects of multiple types of interventions without being limited to a specific form of algorithmic action.

\begin{table*}[]
\small
\setlength{\tabcolsep}{4.5pt}
\centering
\begin{tabular}{lcrrrrrrr}
\toprule
{\bf Dataset} & {\bf Transformation} &
\textbf{\#Posts} & \textbf{\#Replies} & \textbf{\#Retweets} & \textbf{\#Quotes} & \textbf{\#Accounts} & \textbf{\#Links} & \textbf{\#Domains}  \\

  \midrule
 US Elections & 
 Main Dataset -- All Content &
  34,762,696 &
  23,906,163 &
  - &
  3,883,501 &
  3,844,610 &
  2,320,840 &
  75,876 \\ 
Ukraine-Russia  & with Valid View Count & 
  4,306,644 &
  2,208,374 &
  - &
  252,849 &
  1,509,877 &
  631,960 &
  53,975 \\   \midrule

US Elections & Network \& Latent & 
  - &
  4,457,343 &
  - &
  - &
  493,149 &
  - &
  - \\ 
Ukraine-Russia  & Ideology Estimation & 
  - &
  - &
  9,939,787 &
  - &
  67,926 &
  - &
  - \\   \midrule

US Elections &   Claim \& & 
  2,451,532 &
  - &
  - &
  - &
  652,851 &
  - &
  - \\ 
Ukraine-Russia  &Topic Tracking & 
  1,845,285 &
  - &
  - &
  - &
  583,803 &
  - &
  - \\ 
  \bottomrule
\end{tabular}
\caption{\revision{C5}{Summary statistics of the dataset used in the analysis. Retweets have likes, replies, and quotes counts equal to zero, and they inherit retweet and view counts from the original tweets. Therefore, they have been used only to compute ideology in the Ukraine-Russia War dataset and excluded from the rest of the analysis.}}
\label{tab:sumstat}
\end{table*}

\section{Dataset}\label{sec:dataset}
We now present the datasets analyzed in our study, specifically, two sets of Twitter/X posts that include view counts, i.e., the total number of times they have been viewed.

\descr{Ukraine-Russia War Dataset.} In 2023, Baqir et al.~\cite{baqir2025unveiling} collected tweets related to military aid in the Ukraine-Russia war through the Academic API.\footnote{The dataset is available at \url{https://osf.io/5m3vr/}.}
More precisely, they used the following set of keywords: `military aid,' `military support,' `tanks,' `Abrams,' `Leopard,' `Challenger,' `jet,' `aircraft,' `munitions,' `HIMARS,' `rockets,' and `missile.'
In addition to tweet metadata typically found in other datasets, this dataset also includes view counts for each piece of content.

Overall, the Ukraine-Russia war dataset includes over 17 million tweets from more than 5.2 million users, posted between November 22, 2022, and March 1, 2023, and collected in April 2023. 
The temporal gap between tweet creation and collection ensures that engagement and view metrics have stabilized, thus enabling robust analysis~\cite{pfeffer2023half} and providing a meaningful snapshot of the discourse surrounding the Ukraine-Russia conflict on Twitter~\cite{baqir2025unveiling}.

\descr{2024 US Elections.} The second dataset was collected by Balasubramanian et al.~\cite{balasubramanian2024public} using a custom-built scraper.\footnote{The dataset is available at \url{https://github.com/sinking8/usc-x-24-us-election}.}
The authors curated a list of 44 keywords related to the 2024 US Elections, including `2024 Elections,' `Donald Trump,' `letsgobrandon,' `ultramaga,' etc.
Following retrieval, we filter the data as discussed in Section~\ref{sec:preproc} and obtain approximately 35 million tweets with valid view counts by 3.8 million users.

\smallskip Arguably, these two datasets are well-suited to the purpose of our analysis, i.e., studying the occurrence and characteristics of shadow banning.
More precisely, the topic of the Ukraine-Russia war has polarized public opinion into supporters and opponents with clearly distinct stances~\cite{la2023retrieving}. This allows us to identify the factions involved in the debate and evaluate whether Twitter/X has been altering the visibility of specific narratives. 
Similarly, discourse around US Elections is inherently polarized; thus, Balasubramanian's dataset~\cite{balasubramanian2024public} lends itself to identifying opposing factions (i.e., supporters of different political candidates) and enables the analysis of visibility alterations at the community level.

\revision{C5}{As we apply several transformations to the datasets throughout the analysis (see Section~\ref{sec:processing}), we provide statistics for the datasets used in the main steps of this study in Table~\ref{tab:sumstat}.}

\section{Methods}\label{sec:processing}
For both datasets, we pre-process the data to filter out possible inconsistencies in the view count statistics.
We also merge content domain information with third-party data to assign political leaning and factuality labels to tweets.
Finally, we select a set of the most prominent accounts, referred to as ``influencers,'' representing the two opposing factions of each debate.
The influencers are then used to infer users' ideological stances and analyze visibility variations with respect to ideology.
In this section, we discuss these actions in detail.

\subsection{Pre-Processing}\label{sec:preproc}
For a sample of tweets, we manually verify that the view counts in our datasets are broadly consistent with those displayed on the Twitter/X platform.
While we observe minor discrepancies, these are likely due to the interval between data collection and manual verification, during which view counts may have slightly changed. 
Overall, our sanity check confirms the reasonable reliability of the datasets from~\cite{baqir2025unveiling,balasubramanian2024public}.

We also perform a filtering procedure to remove potentially inconsistent records, specifically, content for which the view count is unavailable and content for which the number of interactions (i.e., likes, comments, retweets, or quotes) exceeds the recorded number of views, as users must view the content before engaging with it.

\subsection{Factuality and Political Bias Estimation}
\label{sec:bias}
To test the dependence of visibility variations on content-level features like political bias and factuality (cf.~RQ1 in Section~\ref{sec:intro}), we rely on third-party classifications to infer the political bias and factuality of the news sources referenced in the debate.
More precisely, we use a dataset obtained from Media Bias/Fact Check (MBFC), an independent fact-checking organization that categorizes news outlets based on their factuality and political bias.\footnote{\url{https://mediabiasfactcheck.com/}.} 
This includes 2,190 different news outlets, along with their domain names, political leaning, and factuality, and classifies outlets across the political spectrum, from ``extreme left'' to ``extreme right.'' 
Additionally, some outlets are labeled as ``questionable'' or ``conspiracy-pseudoscience'' if they frequently publish misinformation, false content, or endorse conspiracy theories. 

We use the MBFC dataset to assign factuality (from `Very High' to `Very Low') and political bias (`Extreme Right' to `Extreme Left') to tweets based on the domain referenced. 
For example, a tweet linking to a CNN article is classified as `Mostly Factual' with a `Left-Center' bias, and one linking to RT News has a `Very Low' factuality with a right-center bias.

\subsection{Influencers Selection}
\label{sec:influencers}
Studying the relationship between variations in visibility and individual features like ideological stance (cf.~RQ2) requires us to infer users' opinions. %
To this end, we use Latent Ideology Estimation~\cite{flamino2023political}, a flexible and powerful method for inferring users' ideological stances that relies on the identification of a set of influential accounts actively engaged in the debate.
This group of users, referred to as influencers, must encompass several subcategories and represent a broad range of opinions across the ideological spectrum. 

For the Ukraine-Russia dataset, we follow the procedure explained in~\cite{baqir2025unveiling} and build a network using retweet interactions, ranking accounts based on their in-degree, i.e., the number of unique users who retweeted them.
We then manually select a set of prominent accounts representing both sides of the debate, prioritizing those with the highest in-degree.
This initial set serves as a seed, and we expand it by utilizing Twitter's ``Who to follow'' recommendations from the selected accounts' profiles. 
We repeat this process until no new relevant accounts are suggested. 
The selection is then refined by excluding accounts with an in-degree below 100 or those whose content was unrelated to the Ukrainian conflict. 
Ultimately, this yields a final set of 204 influencers, representing both supporters and opponents of military support to Ukraine.

For the 2024 US Elections dataset, retweet interactions are not sufficiently rich to enable a reliable estimation of users' opinions, \revision{C6}{as the dataset contains only a few hundred retweets}. 
Therefore, we use reply interactions to reconstruct the interaction network, \revision{C6}{ which, although noisier, still allows us to distinguish between the two groups}. 
We identify the most replied-to users and manually extract the top 100 accounts exhibiting a clear stance, either supporting or opposing Trump. 
This allows us to select a balanced set of influencers representing both pro- and anti-Trump positions. 
Using Hartigan's dip test of unimodality~\cite{hartigan1985dip}, we confirm the bimodal nature of the latent ideological estimation distribution for users and influencers \revision{}{in both datasets (Ukraine-Russia War: D = 0.027, p-value $< 2.2e-16$ for users and D = 0.074, p-value $= 5.078e-06$ for influencers; US Elections: D = 0.012, p-value $< 2.2e-16$ for users and D = 0.106, p-value $< 2.2e-16$ for influencers)}. 
This result highlights the polarized nature of the debate, as well as the reliability of our inference~\cite{flamino2023political}.

\subsection{Users' Ideological Stance Estimation}
\label{sec:latent}
Latent Ideology Estimation is a proven method for reliably estimating user ideology in diverse contexts~\cite{flamino2023political,barbera2015tweeting, falkenberg2022growing}. 
Using the set of accounts identified using the methodology described above, we apply this algorithm to the Ukraine-Russia war and the US 2024 Elections debates to infer users' stances and contribute to answering RQ2.
Crucially, this algorithm depends on identifying a subset of key influencers, as their selection significantly affects the accuracy of the ideology estimation.

Once the influencer set is identified, we apply the Correspondence Analysis algorithm~\cite{greenacre2010correspondence}, performing three main steps: 1) constructing an interaction matrix, 2) normalizing it, and 3) performing singular value decomposition (SVD). 
More precisely, in the first step, we build the matrix $A$, where each element $A_{ij}$ represents the number of interactions (retweets for Ukraine-Russia, replies for US 2024 elections) that user $i$ directs to influencer $j$. We then normalize $A$ by dividing it by the total number of interactions, yielding $P = \frac{A}{\sum_{ij} A_{ij}}$.
Using the following quantities: $\textbf{r} = P \textbf{1}$, $\textbf{c} = \textbf{1}^T P$, $D_r = \text{diag}(\textbf{r})$, and $D_c = \text{diag}(\textbf{c})$, we normalize further to obtain $S = D_r^{-1/2}(P-\textbf{r}\textbf{c})D_c^{-1/2}$.
In the third step, we perform singular value decomposition of $S$, such that $S = U \Sigma V^T$, where $U$ and $V$ are orthogonal matrices and $\Sigma$ is a diagonal matrix of singular values. 
We estimate user ideological leaning by taking the subspace associated with the first singular value of the decomposition. 
Specifically, the latent ideology of user $i$ is the $i$-th entry in the first column of the matrix $U$, while the ideology of each influencer is estimated as the median ideology score of their retweeters/repliers.

\revision{\descr{[C6] Retweets vs. Replies.} For the US Elections dataset, we use replies instead of retweets since the researchers collecting US Elections datasets were only able to collect a few hundred retweets, which is insufficient to run latent ideology estimation for inferring users' leanings.
While replies can be a bit noisier, they can still separate users based on their political stance.}

\subsection{Claim Detection and Tracking}
\label{sec:claim_tracking}
To get a more nuanced understanding of the different types of textual content and visibility patterns associated with them, we focus on claim-level analysis of tweets.
This is an additional vantage point to help us answer RQ1, with an emphasis on the claims being spread around different topics.
To this end, we use Natural Language Processing (NLP) and Learning to Rank (LTR) techniques to group tweets into a coherent structure of claims.

First, we use the claim extraction pipeline from Lambretta~\cite{paudel2023lambretta} to extract claims from tweets that make an objective claim and are candidates for being tracked through downstream keyword extraction or semantic search methodologies.
These claims are structurally similar to those manually curated by fact-checking organizations like Snopes or PolitiFact.\footnote{See \url{www.snopes.com}, \url{https://www.politifact.com/}}.
We sample 1,000 unique tweets sorted by several different criteria of engagement: 1)~Retweet count, 2)~Reply count, 3)~Like count, 4)~Quote count, and 5)~\revision{}{Views} count.
This sampling gives us a set of tweets that have a higher likelihood of containing a claim (and thus garnered the level of interaction on Twitter), rather than sampling tweets randomly.
We then use ClaimSpotter~\cite{hassan2017toward} to identify tweets that contain a claim: ClaimSpotter returns a score between 0 and 1, representing how ``check-worthy'' an input text is.
Consistent with~\cite{paudel2023lambretta}, we select 0.5 as the threshold for classifying tweets containing a claim.
For the filtered tweets containing a claim, we use Lambretta's keyword identification component to extract the best set of keywords representing the claim. 
Finally, we retrieve similar tweets that make the same claim as the seed set using the extracted set of keywords and use these subsets of tweets to analyze different visibility patterns.

\begin{figure*}[t]
    \centering
    \begin{subfigure}[t]{0.45\linewidth}\vskip 0pt
    \includegraphics[width=\linewidth]{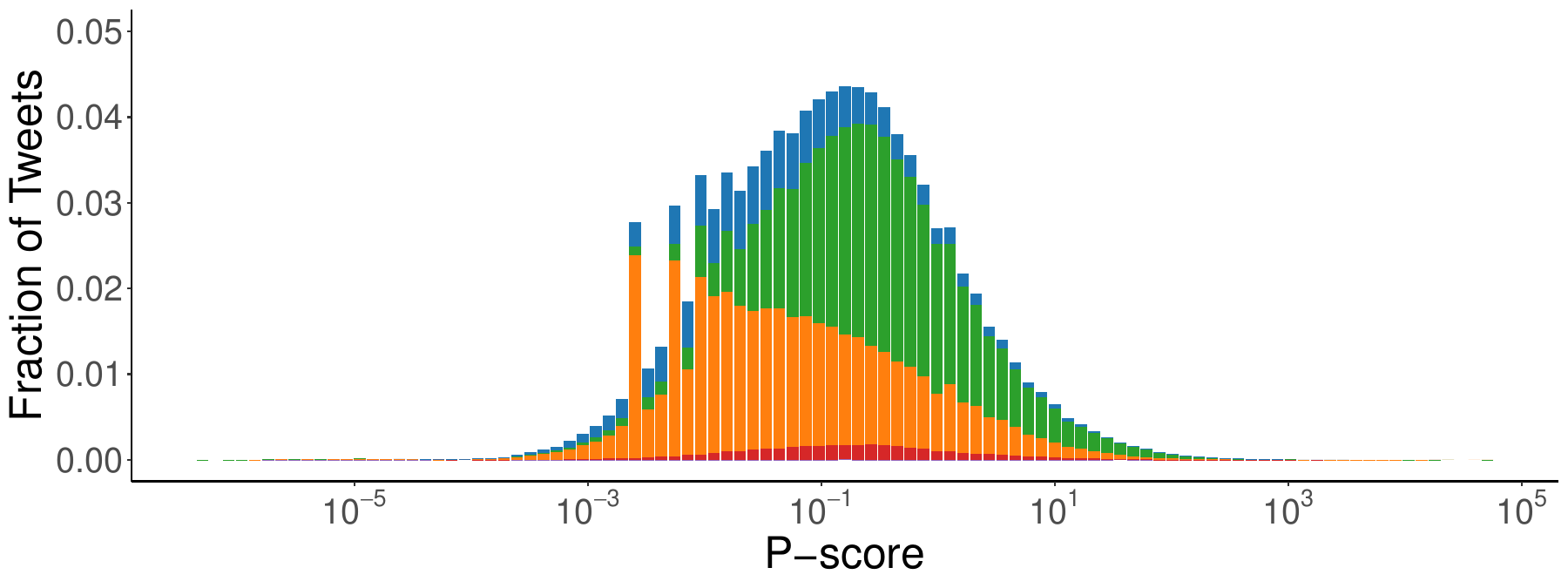}
    \caption{Ukraine-Russia War}
    \end{subfigure}
    ~
    \begin{subfigure}[t]{0.45\linewidth}\vskip 0pt
    \includegraphics[width=\linewidth]{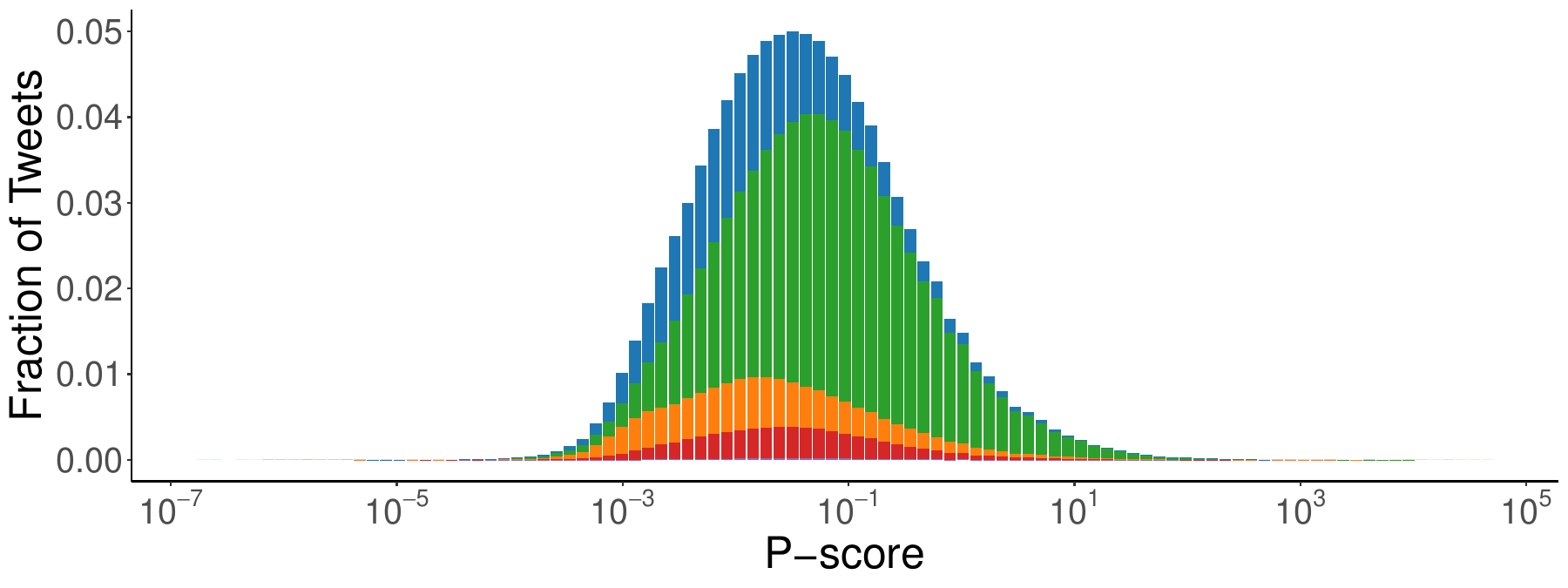}
    \caption{2024 US Elections}
    \end{subfigure}
    \hspace{-1.5cm}
    \begin{subfigure}[t]{0.09\linewidth}\vskip 6pt
    \includegraphics[width=\linewidth]{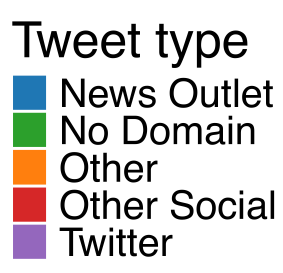}
    \end{subfigure}
    \caption{P-score distributions of original content for the Ukraine-Russia War and the 2024 US Elections datasets. The color represents the different categories of tweets.}
    \label{fig:p_score_all_content}
\end{figure*}

Since the tweets in the datasets span longitudinally and contain a multitude of events (both related to the Ukraine-Russia war and the 2024 US Elections), we further label the claims identified in the seed set of tweets to different themes.
This makes the analysis tractable while allowing us to achieve the granular content level analysis.
We use the unsupervised topic modeling library BertTopic~\cite{grootendorst2022bertopic} to identify coherent topics within the seed set of claims.
We select a laxer threshold (number of documents in a topic, and similarity measure for two claims to be assigned the same topic), resulting in a larger number of initial topics.
These topics are then manually cleaned for noise (artifacts of keywords used in the initial data sampling process) and iteratively assigned to themes.
This way, we have a hierarchy of themes within the seed set of claims and a subset of tweets across themes whose visibility patterns can be compared.

\subsection{P-score}
As mentioned earlier, we use the ``view count'' metric to examine differences in visibility across the Ukraine-Russia War and the 2024 US Elections datasets.
We do so as shadow banning essentially is the act of reducing the visibility of specific profiles or content through various techniques~\cite{le2021setting,gillespie2022not}.
Thus, view counts are an appropriate proxy for quantifying visibility. 
However, directly comparing the visibility of users may prompt some challenges -- more precisely, content visibility on social networks heavily depends on the popularity of the author, as social media algorithms use follower-following relationships to suggest content~\cite{vombatkere2024tiktok}.

As a result, we introduce and use a metric denoted as \textit{p-score}, which accounts for the author's popularity as follows: 
\begin{equation}
\text{p-score}= \frac{\text{view count}}{\text{number of followers}}
\end{equation}

In other words, the p-score of a tweet quantifies the ratio between the number of views it receives and the number of followers of its author. 
Hence, a higher p-score indicates that content receives more views per follower, while lower values suggest more limited circulation.

\section{Detecting Visibility Variations}

In this section, we set out to answer our three research questions (see~Section~\ref{sec:intro}) related to studying the occurrence of shadow banning at content, user, and network levels.

\begin{figure*}[t!]
    \centering
    \begin{subfigure}[t]{0.44\linewidth}\vskip 0pt
    \includegraphics[width=\linewidth]{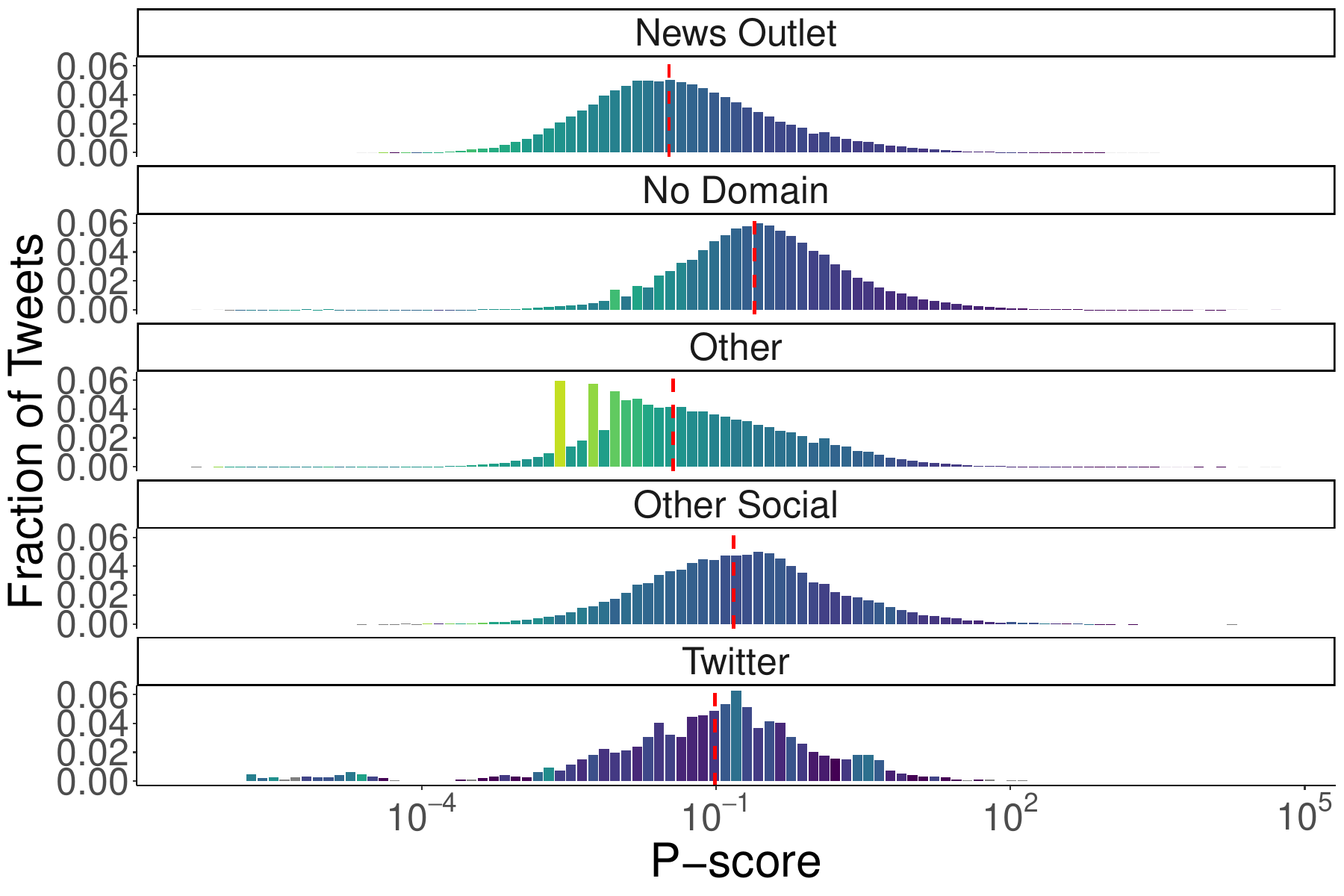}
    \caption{Ukraine-Russia War}
    \end{subfigure}
    ~
    \begin{subfigure}[t]{0.44\linewidth}\vskip 0pt
    \includegraphics[width=\linewidth]{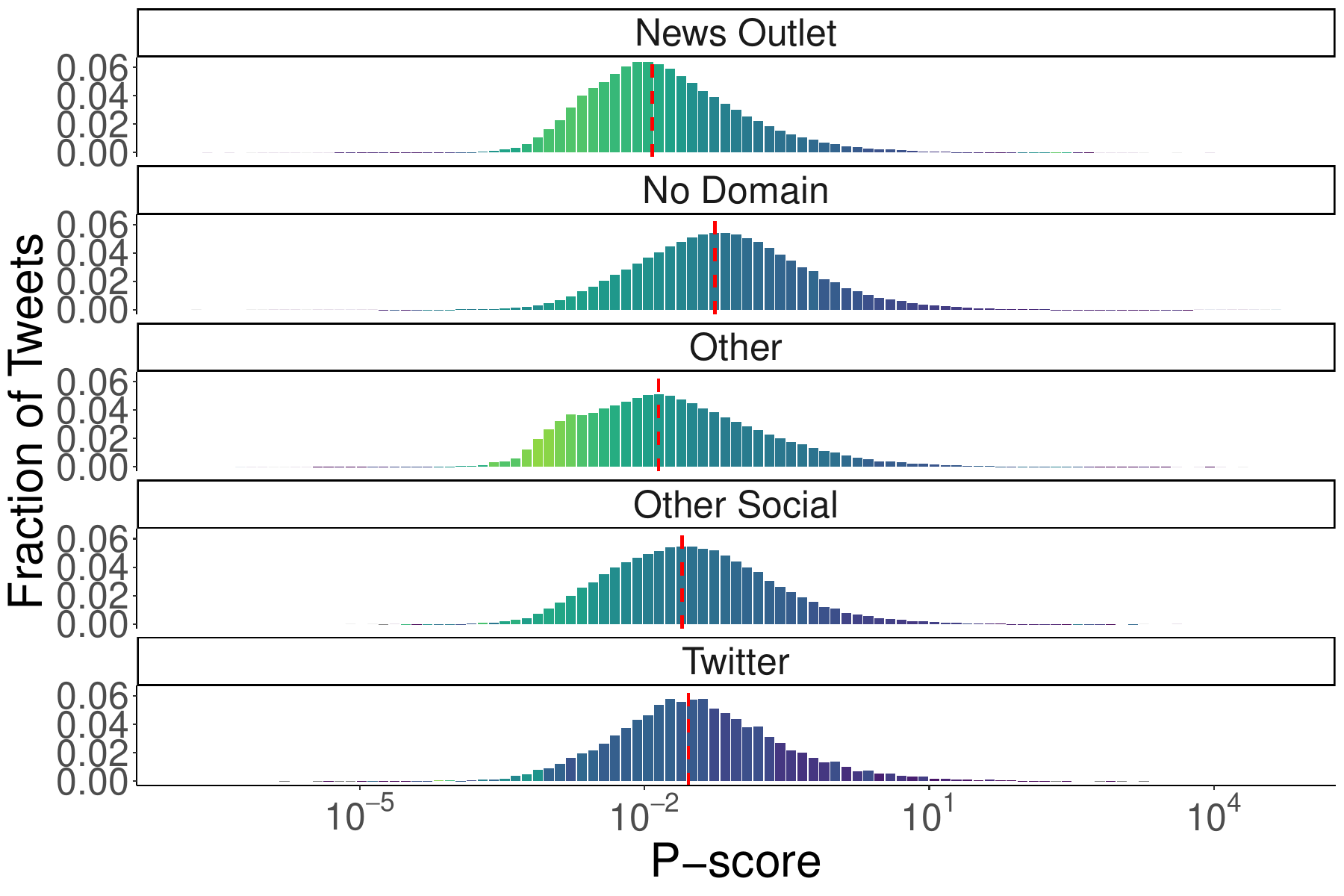}
    \caption{2024 US Elections}
    \end{subfigure}
    \begin{subfigure}[t]{0.09\linewidth}\vskip 8pt
    \includegraphics[width=\linewidth]{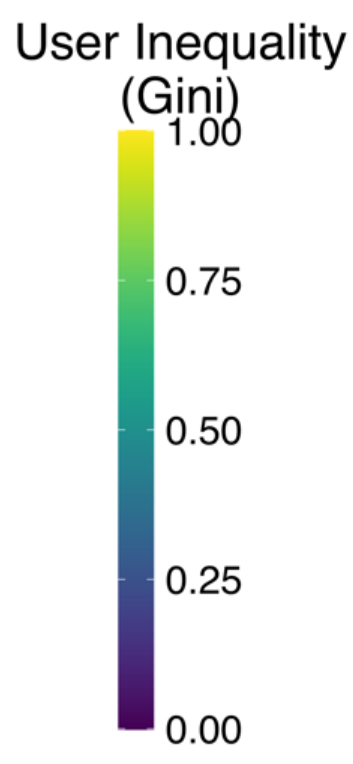}
    \end{subfigure}
    \caption{P-score distributions by tweet category. The colors represent the disparity in users' contributions to each interval as per the Gini index. Higher values indicate that a small number of users generate most of the content within the interval, while lower values suggest a more uniform distribution. Red lines mark the median of the distribution for each category.}
    \label{fig:disparity_user_pscore}
\end{figure*}

\subsection{RQ1: Content Level}

To analyze the occurrence of shadow banning at the content level, we examine whether the p-scores of the tweets depend on the type of information they convey. 
To ensure a fair comparison and avoid possible confounding factors, we exclude replies, retweets, and quotes from the analysis, ending up with 1.8M tweets for the Ukraine-Russia dataset and 7.2M tweets for the 2024 US Elections dataset. 

We classify tweets based on the domain they refer to using the MBFC dataset (see Section~\ref{sec:bias}), dividing them into five categories: ``News Outlets,'' ``No Domain,'' ``Other,'' ``Other Social,'' and ``Twitter.''
These refer, respectively, to tweets that contain a link to a news outlet, no link, a link to an unclassified website, a link to another social platform, and a link to other Twitter content.

\descr{Content P-Score Distributions.} 
We begin by examining whether tweets, categorized by the type or absence of external URLs, exhibit differences in p-score distributions.
In Figure~\ref{fig:p_score_all_content}, we plot the distribution of the p-scores for all original content (i.e., without considering replies, retweets, and quotes) for each category of content. 
For both datasets, the distribution has a log-normal shape, even though the categories contribute differently to the distribution density. 
For instance, the ``Other'' and ``News Outlets'' categories contribute more to lower values, while ``No Domain'' appears to be skewed toward higher values.
Also, the spectrum of this distribution can vary widely, spanning nine orders of magnitude. 

Although the wide range of p-scores can be explained by the typical dynamics of social networks, whereby some content goes viral, some is ignored, and the majority receives a few interactions, the spectrum's unbalanced and asymmetric constitution suggests the presence of different visibility levels for different categories.
Tweets that do not link to other domains seem favored over others in both datasets.
However, this difference seems to be considerably more pronounced in the Ukraine-Russia dataset than for the 2024 US Elections. 
This may suggest a varying degree of intervention over time, while also highlighting the algorithm's systematic promotion of content without external URLs.

\descr{Artifacts in Content Visibility Distributions.} We also observe the presence of peaks in certain intervals of the distributions in the Ukraine-Russia dataset.
This may depend on the systematic penalization of specific users or the algorithm detecting inappropriate content and limiting its circulation. 
We analyze the distribution of user contributions within each interval to better understand the causes of these outliers and the role of users in shaping them. 
To do so, we use the Gini index, a common metric to measure the inequality of distributions, computing it on the distribution of the number of tweets per interval for each author~\cite{yitzhaki2013gini}.
In our case, if all users contribute to the population of an interval with approximately the same number of tweets, the Gini index is close to 0, whereas if a significant disparity exists, with some users generating most tweets in the interval, the index approaches 1. 

In Figure~\ref{fig:disparity_user_pscore}, we present the distribution of p-scores across tweet categories for both datasets \revision{C5}{(also see Table~\ref{tab:disparity_user_pscore})}. 
In both cases, the median p-score (marked by the red line) for the ``No Domain'' category is higher than for all others, confirming observations from Figure~\ref{fig:p_score_all_content}.
Specifically, tweets from news outlets exhibit the lowest median p-scores (0.033 for Ukraine-Russia, 0.012 for US Elections), while ``No Domain'' content shows significantly higher medians (Ukraine-Russia: 0.246; US Elections: 0.056). 
This further supports the hypothesis that Twitter's algorithm favors content that does not contain external URLs.
Interestingly, tweets linking to unclassified websites have median p-scores close to those of news outlets (Ukraine-Russia: 0.036; US Elections: 0.014), whereas tweets quoting other tweets or linking to other social platforms are slightly less penalized (median p-scores of 0.097 and 0.150 for Ukraine-Russia and 0.029 and 0.024 for US Elections).
To validate these observations, we perform a Mann-Whitney U test between all pairs of content categories to assess whether the values in one group tend to be greater than those in another~\cite{mann1947test}. 
The Mann-Whitney U test does not assume normality and instead evaluates whether there is a statistically significant shift between distributions. 
In our case, a low p-value indicates strong evidence that the values from one group are systematically higher than those from another. 
For both the Ukraine-Russia and US Elections datasets, the tests confirm that the ``No Domain'' category statistically significantly dominates all others (p-value $< 2.2e-16$ for all comparisons).

\begin{table}[t!]
\centering
\smallskip
\small
\begin{tabular}{lrr}
\toprule
\textbf{Content} & \textbf{Ukraine-Russia} & \textbf{US Elections} \\ \midrule
{News Outlet}   & 252,046                  & 1,564,923               \\ 
{No Domain}     & 796,673                  & 4,317,988               \\ 
{Other}         & 730,078                  & 851,246                \\ 
{Other Social~}  & 65,025                   & 483,346                \\ 
{Twitter}       & 1,463                    & 24,509                 \\ 
\bottomrule
\end{tabular}
\caption{\revision{C5}{Tweets by content type for  distributions from Figure~\ref{fig:disparity_user_pscore}}.}
\label{tab:disparity_user_pscore}
\end{table}

\begin{figure*}[t!]
    \centering
    \begin{subfigure}[t]{0.22\linewidth}\vskip 0pt
        \centering
\includegraphics[width=\linewidth]{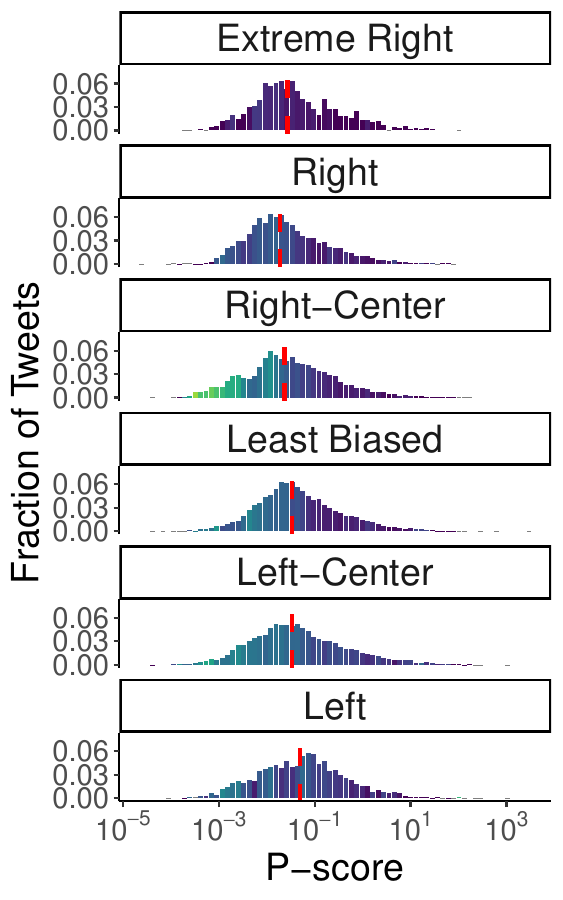}
    \caption{Political bias (Ukraine-Russia)}    \label{bias-a}
    \end{subfigure}
    \begin{subfigure}[t]{0.22\linewidth}\vskip 0pt
    \centering
    \includegraphics[width=\linewidth]{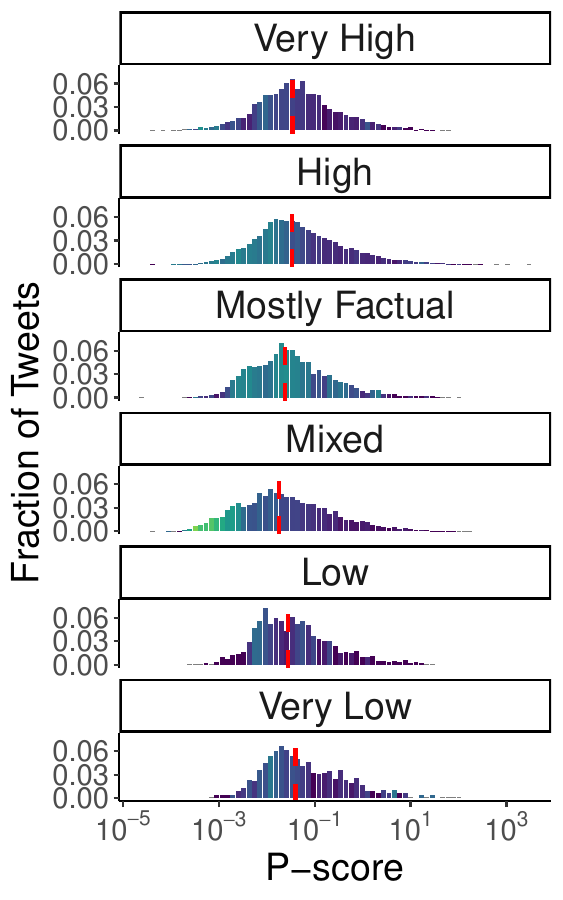}
    \caption{Factuality (Ukraine-Russia)}    \label{bias-b}
    \end{subfigure}
   \begin{subfigure}[t]{0.22\linewidth}\vskip 0pt
    \centering
    \includegraphics[width=\linewidth]{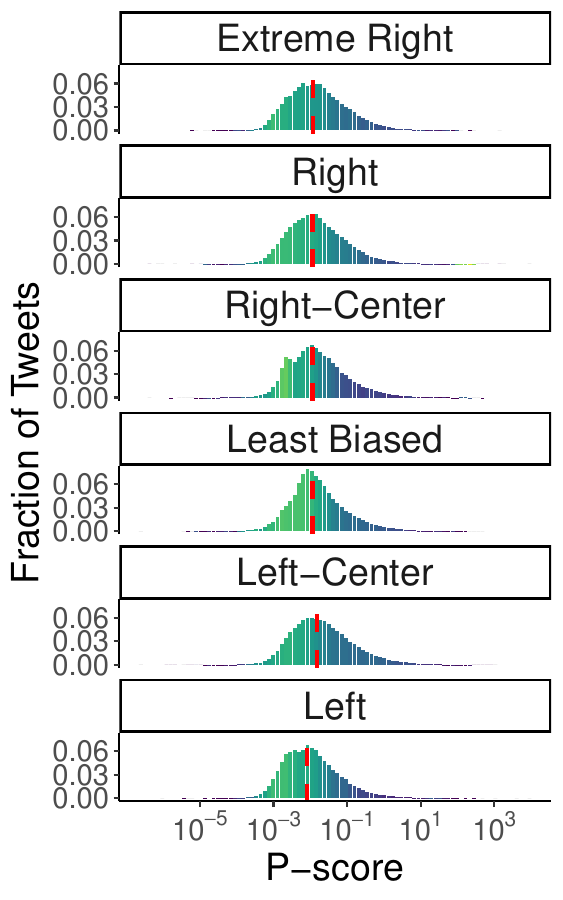}
    \caption{Political Bias (US Elections)} \label{bias-c}
    \end{subfigure}
    \begin{subfigure}[t]{0.22\linewidth}\vskip 0pt
    \centering
    \includegraphics[width=\linewidth]{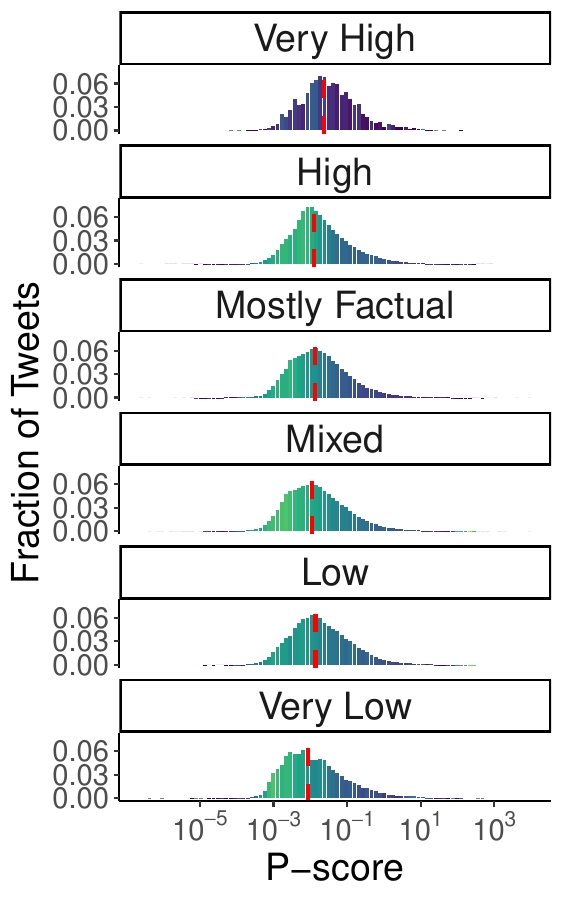}
    \caption{Factuality (US Elections)} \label{bias-d}
    \end{subfigure}
    \begin{subfigure}[t]{0.09\linewidth}\vskip 8pt
    \includegraphics[width=\linewidth]{figures/content_type_gini-legend.pdf}
    \end{subfigure}
    \caption{P-score distributions wrt the political leaning and factuality of the news sources as classified by MediaBias/FactCheck. Red lines mark the median value for each category. Colors represent the disparity in users' contributions to each interval as per the Gini index.}
    \label{fig:reliability_political}
\end{figure*}

For the Ukraine-Russia dataset, the peaks in the ``Other'' distribution correspond to the intervals with the highest Gini scores, indicating the presence of users with a high volume of tweets who receive a similar level of visibility. 
Indeed, most of the content populating these intervals belongs to a few accounts explicitly labeled as automated by the platform, suggesting that the algorithm may be limiting the visibility of their content.
A similar behavior, although less pronounced, emerges in the 2024 US Elections dataset.
The highest level of imbalance appears in the left tail of the ``Other'' category distribution, where a substantial portion of tweets comes from a few highly active accounts that primarily discuss news by frequently linking to external websites.
For instance, the most represented account in the bins with the highest disparity values has an average activity of 38 tweets per day, with 97.5\% of its content linking to external websites, suggesting that the algorithm may further penalize very active users who frequently post URLs.

Overall, these artifacts in the distributions of content visibility reflect an unbalanced contribution by users for both datasets, pointing to interventions aimed at reducing the visibility of specific accounts.
These patterns also confirm the consistent application of such practices over time.

\begin{table*}[t!]
\centering
\small
\begin{subtable}[t]{0.45\textwidth}
\centering
\begin{tabular}{lrr}
\toprule
\textbf{Bias}              & \textbf{Ukraine-Russia} & \textbf{US Elections} \\ 
\midrule
{Extreme Right~} & 929                     & 124,547                \\ 
{Right}         & 4,211                    & 427,986                \\ 
{Right-Center}  & 9,586                    & 165,889                \\ 
{Least Biased}  & 9,838                    & 200,011                \\ 
{Left-Center}   & 11,935                   & 484,472                \\ 
{Left}          & 3,204                    & 148,036                \\ 
\bottomrule
\end{tabular}
\caption{Political Bias}
\end{subtable}
\begin{subtable}[t]{0.45\textwidth}
\centering
\begin{tabular}{lrr}
\toprule
\textbf{Factuality}               & \textbf{Ukraine-Russia} & \textbf{US Elections} \\ 
\midrule
{Very Low}       & 845                     & 46,747                 \\ 
{Low}            & 940                     & 101,638                \\ 
{Mixed}          & 10,485                   & 668,796                \\ 
{Mostly Factual} & 4,068                    & 218,133                \\ 
{High}           & 19,158                   & 513,152                \\ 
{Very High}      & 2,818                    & 3,230                  \\ 
\bottomrule
\end{tabular}
\caption{Factuality}
\end{subtable}
\caption{\revision{C5}{Number of tweets by political bias and factuality for the distributions from Figure~\ref{fig:reliability_political}.}}
\label{tab:reliability_political}
\end{table*}

\begin{figure*}[t]
    \centering
    \begin{subfigure}[t]{0.48\linewidth}\vskip 0pt
    \includegraphics[width=\linewidth]{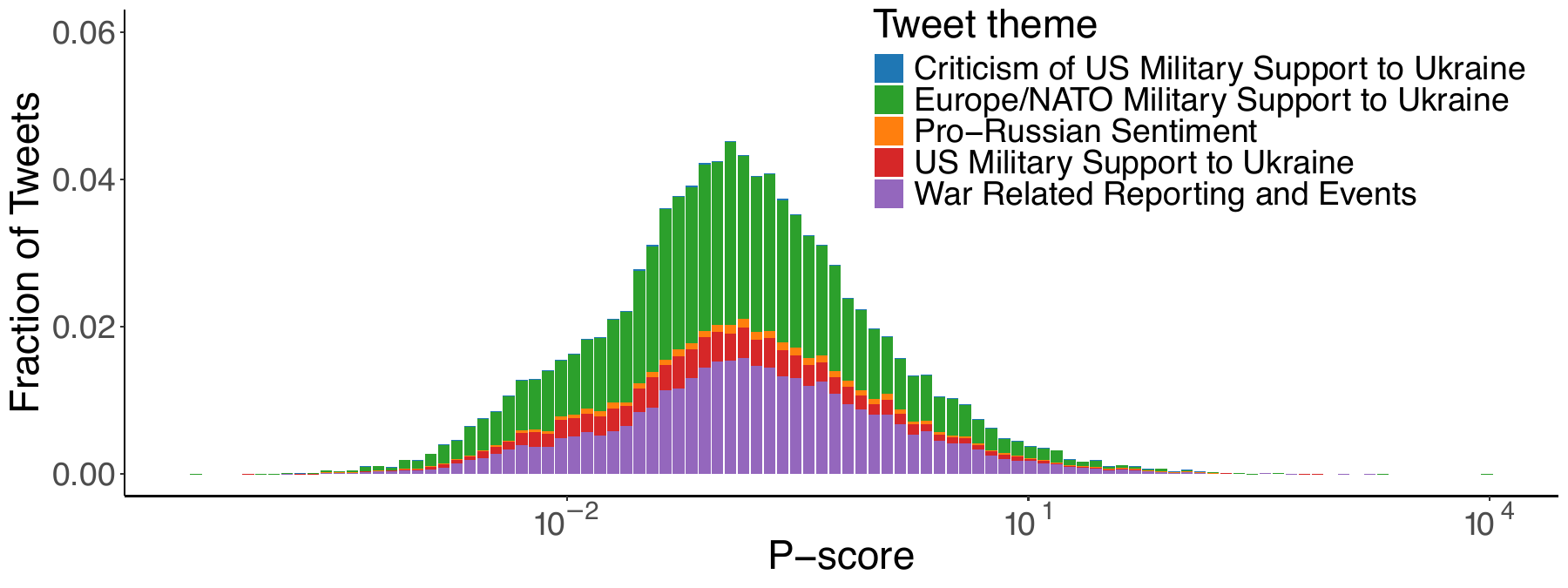}
    \caption{Ukraine-Russia War}
    \end{subfigure}
    ~
    \begin{subfigure}[t]{0.48\linewidth}\vskip 0pt
    \includegraphics[width=\linewidth]{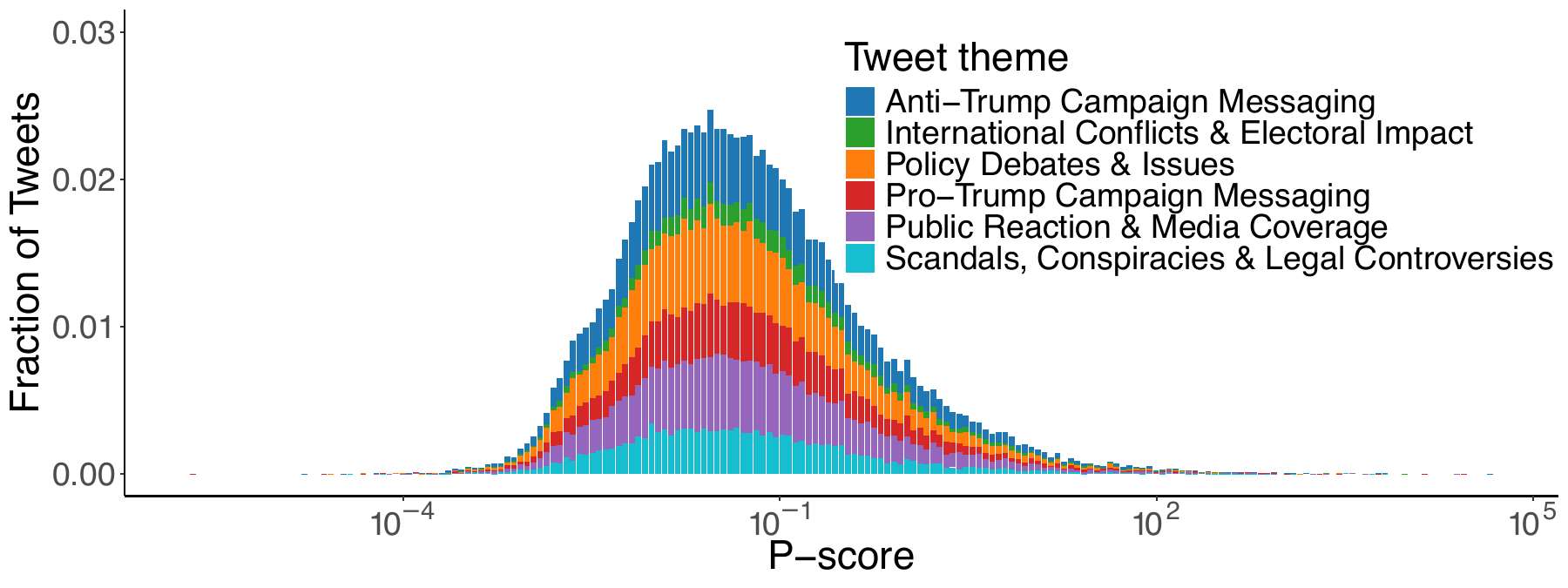}
    \caption{2024 US Elections}
    \end{subfigure}
    \caption{P-score distributions of original content with colors representing the different themes of tweets, highlighting the contributions of each theme depending on the p-score.}
    \label{fig:p_score_lambretta}
\end{figure*}

\descr{Political Bias and Factuality.}
Next, we analyze the dependence of visibility on the political bias and factuality of the content being shared.
We apply the same methodology to assess whether visibility varies based on the political bias and factuality of the sources referenced in tweets, for both the Ukraine-Russia War (Figure~\ref{bias-a}--\ref{bias-b}) and the 2024 US Elections (Figure~\ref{bias-c}--\ref{bias-d}) debates \revision{}{(also see Table~\ref{tab:reliability_political})}.
For both datasets, the results show a relatively contained variation in visibility across both political bias (Ukraine-Russia: min 0.019, max 0.05; US Elections: min 0.0081, max 0.015) and factuality (Ukraine-Russia: min 0.018, max 0.04; US Elections: min 0.0086, max 0.024), suggesting that the algorithm does not systematically penalize content based on these attributes.
This is also confirmed via the Mann-Whitney U test, which did not find a dominant category for political leaning in the Ukraine-Russia dataset (all categories exhibit p-values $>$ 0.05 in at least one comparison).
As for factuality, the tests highlight that the ``Mixed'' and ``Low'' categories tend to be penalized (maximum p-values $< 2.2e-16$ and 0.0019, respectively), but reject the hypothesis that ``Very Low'' content is consistently outperformed by ``High'' or ``Very High'' categories.
Similarly, in the US Elections dataset, there is no evidence of clear dominance among political leanings (all categories have p-values $>$ 0.05 in at least one comparison), although the tests detect the lower performance of the ``Left'' category (p-value $< 2.2e-16$).
Interestingly, for factuality, ``Very High'' content outperforms all other categories (p-values $< 2.2e-16$ in all comparisons), while no clear dominance emerges among the remaining classes.
However, the number of tweets in the ``Very High'' category is considerably smaller than in others (\revision{}{see Table~\ref{tab:reliability_political}}), suggesting an occasional rather than systematic consumption of highly factual content in the US Elections debate.

Notably, the ``Right-Center'' and ``Mixed'' categories in the Ukraine-Russia dataset exhibit a higher Gini index in some lower p-score bins, suggesting the possible presence of penalized domains belonging to those categories.
For the US Elections dataset, this effect is less pronounced, with only the ``Right-Center'' category showing minor anomalies with slightly higher Gini indices.
Additionally, the latter displays greater variation in the Gini index across the distribution, with lower p-score bins having higher Gini values.
However, this behavior appears to affect all categories of political bias and factuality -- except ``Very High,'' which has considerably fewer observations -- and thus does not suggest targeted interventions to limit the visibility of specific categories.
Overall, the analysis of content features and their influence on visibility highlights that content circulation on Twitter is influenced primarily by the presence of external URLs, rather than by the political stance or factuality of the information conveyed, with tweets containing links to external sources systematically penalized by the recommendation algorithm.

\begin{figure*}[t]
    \centering
    \begin{subfigure}[t]{0.44\linewidth}\vskip 0pt
    \includegraphics[width=\linewidth]{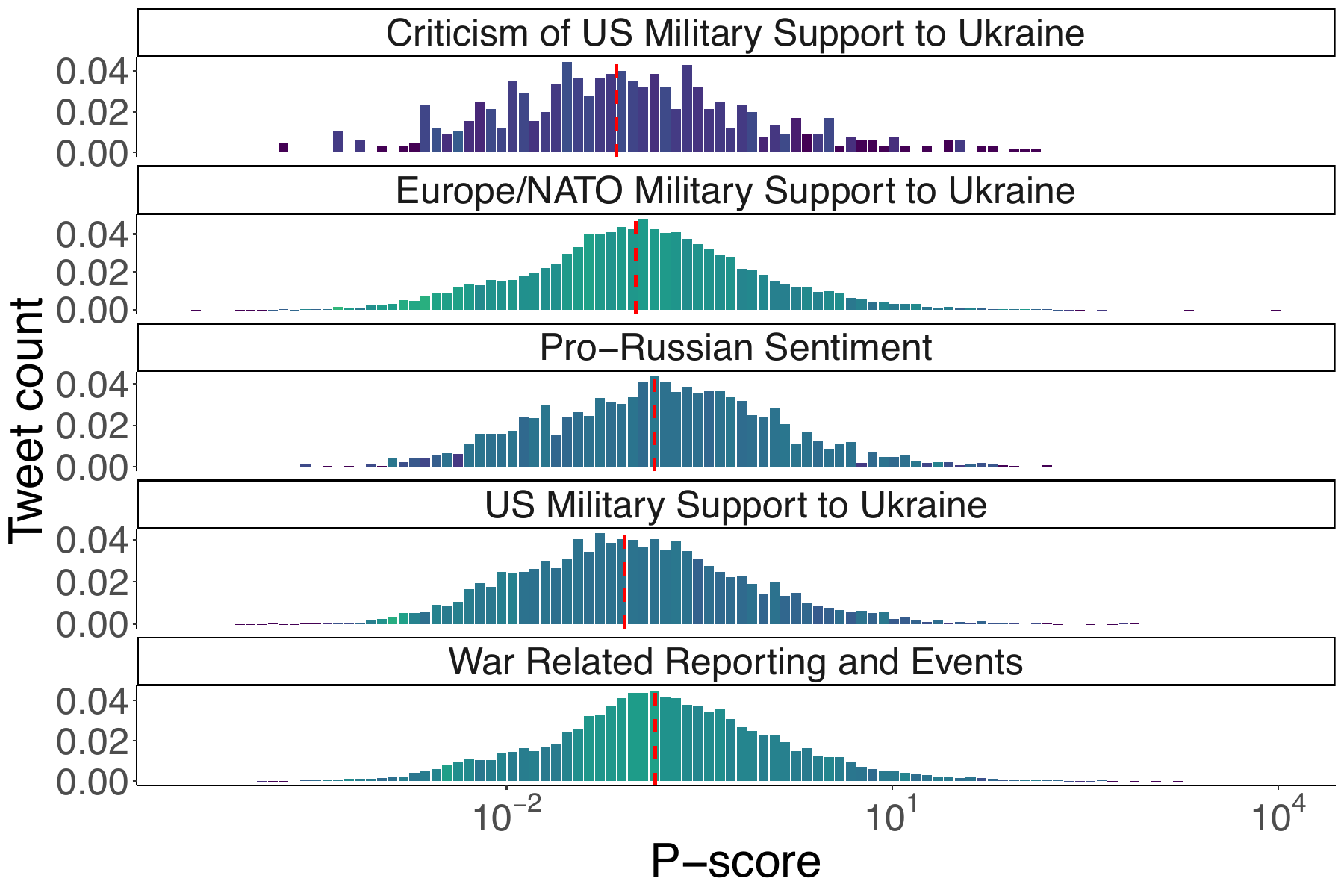}
    \caption{Ukraine-Russia War}
    \end{subfigure}
    ~
    \begin{subfigure}[t]{0.44\linewidth}\vskip 0pt
    \includegraphics[width=\linewidth]{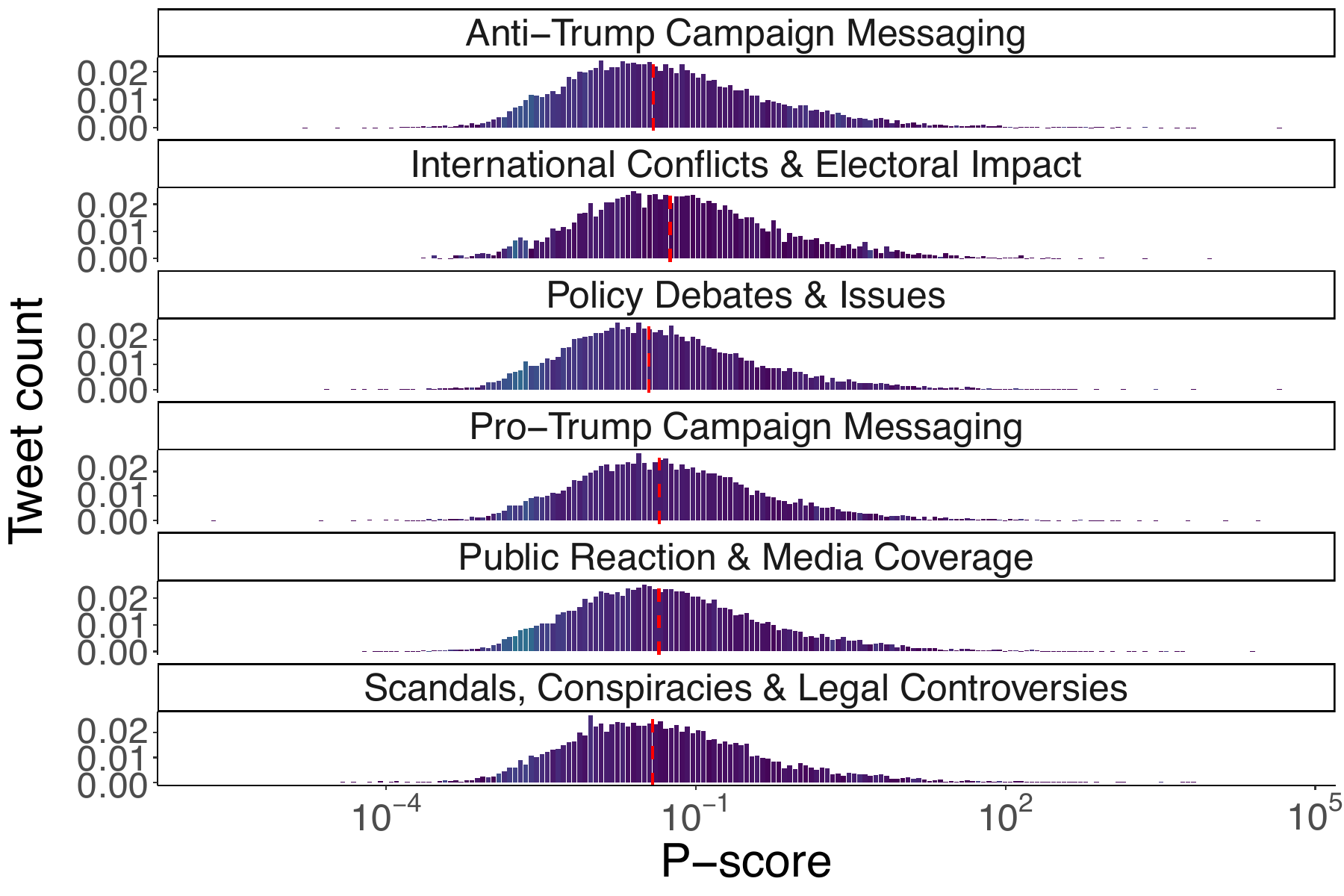}
    \caption{2024 US Elections}
    \end{subfigure}
    \begin{subfigure}[t]{0.09\linewidth}\vskip 8pt
    \includegraphics[width=\linewidth]{figures/content_type_gini-legend.pdf}
    \end{subfigure}
    \caption{P-score distributions by category with colors representing the disparity in users' contributions to each interval as per the Gini index. Higher values indicate that a small number of users generate most of the content within the interval, lower values a more uniform distribution. Red lines mark the median of the distribution for each theme.}
    \label{fig:disparity_lambretta}
\end{figure*}

\descr{Topics.}
Next, we use the textual content of tweets to investigate whether the different topics being discussed experienced varying levels of visibility.
Following the claim detection and tracking methodology discussed in Section~\ref{sec:claim_tracking}, for the Ukraine-Russia dataset, we first filter out 2,735 seed tweets making a claim.
These seed tweets are further categorized into five different topics: 1)~Europe/NATO Military Support to Ukraine, 2)~Criticism of US Military Support to Ukraine, 3)~US Military Support to Ukraine, 4)~War Related Reporting and Events, and 5)~Pro-Russian Support.
Keyword extraction and the subsequent matching procedure return 151,697 tweets discussing claims related to the seed topics.
We applied the same procedure for US elections dataset, categorizing the seeds into the following topics: 1)~Anti-Trump Campaign Messaging, 2)~International Conflicts \& Electoral Impact, 3)~Policy Debates \& Issues, 4)~Pro-Trump Campaign Messaging, 5)~Public Reaction \& Media Coverage, 6)~Scandals, Conspiracies \& Legal Controversies

Figure~\ref{fig:p_score_lambretta} shows the distribution of the p-score values for the claim-matched tweets, divided by each theme of claims. %
Similar to Figure~\ref{fig:p_score_all_content}, the distribution follows a log-normal shape.
However, in both cases the p-scores from the different themes have a largely balanced and symmetric distribution, contributing almost uniformly to the distribution density.

\begin{figure*}[t]
    \centering
    \begin{subfigure}[b]{0.49\linewidth}\vskip 0pt
    \includegraphics[width=\linewidth]{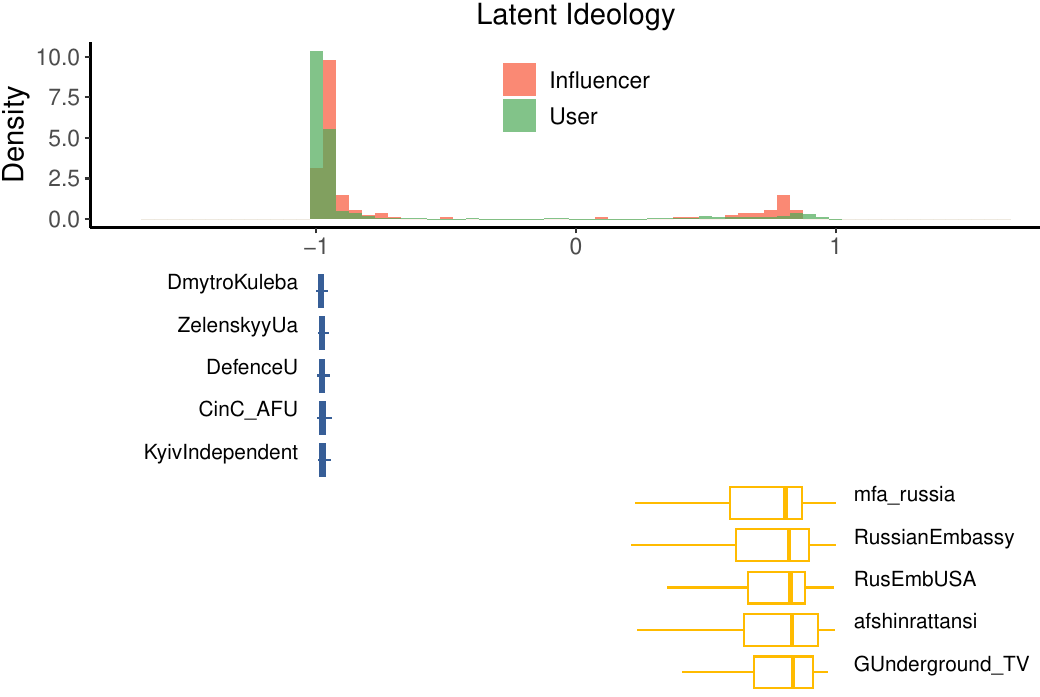}
    \caption{Ukraine-Russia War}
    \end{subfigure}
    ~
    \begin{subfigure}[b]{0.49\linewidth}\vskip 0pt
    \includegraphics[width=\linewidth]{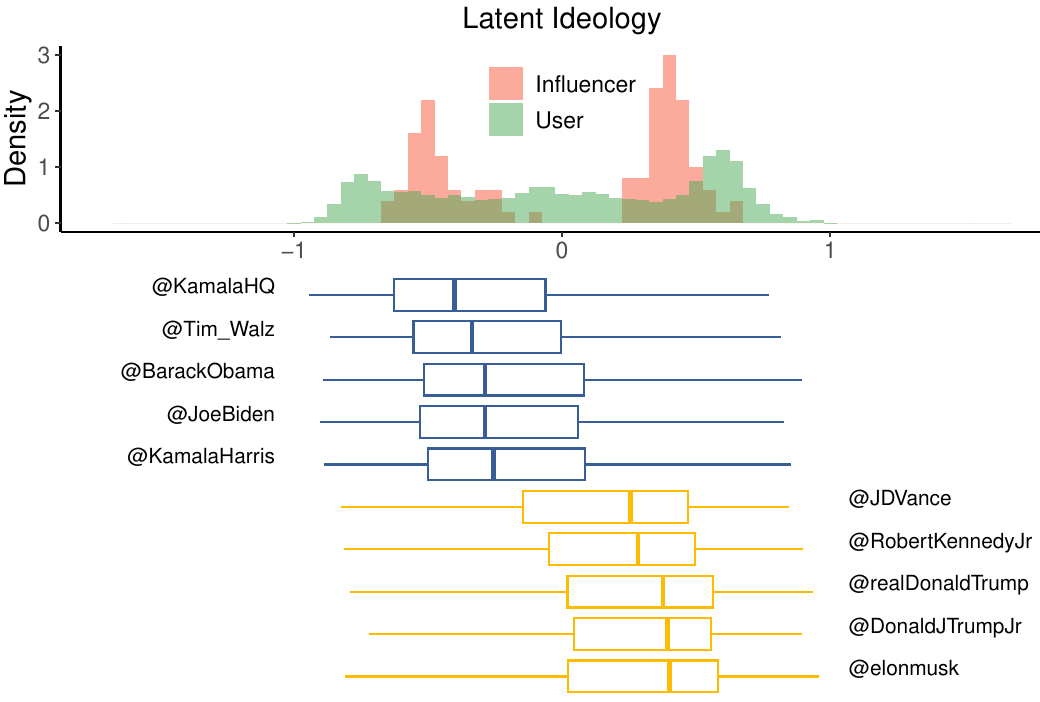}
    \caption{2024 US Elections}
    \end{subfigure}
    \caption{\revision{C6}{Ideology analysis on the two datasets. The bar plots show the distribution of ideology scores for both users and influencers, while the box plots display the distribution of audience ideology scores, defined as retweeters in the Ukraine-Russia War dataset and repliers in the U.S. Elections dataset, for five representative influencers from each faction. A total of 147 prominent accounts are used as influencers for the Ukraine-Russia War dataset, and 100 accounts for the U.S. Elections.}}
    \label{fig:ideology_analysis}
\end{figure*}

\begin{figure*}[t]
    \centering
    \begin{subfigure}[t]{0.49\linewidth}\vskip 0pt
    \includegraphics[width=\linewidth]{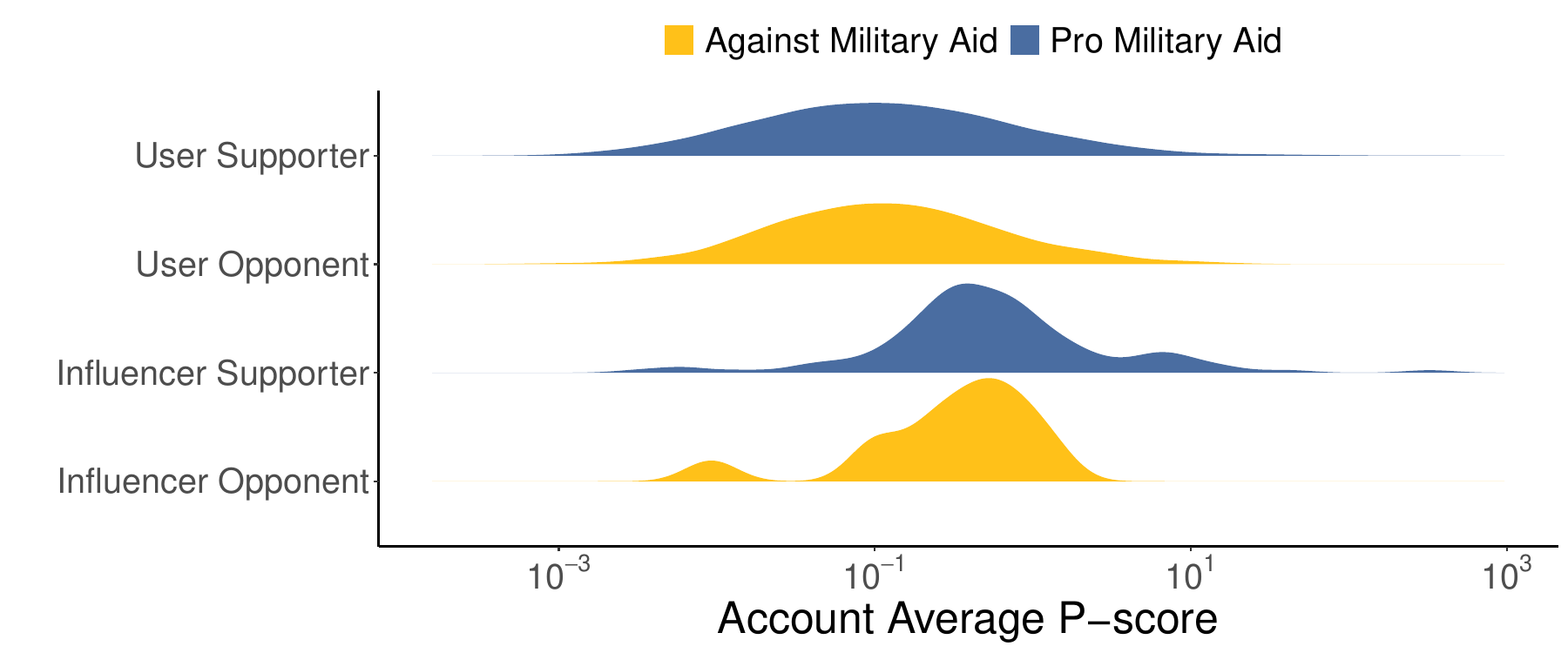}
    \caption{Ukraine-Russia War}
    \end{subfigure}
    \begin{subfigure}[t]{0.49\linewidth}\vskip 0pt
    \includegraphics[width=\linewidth]{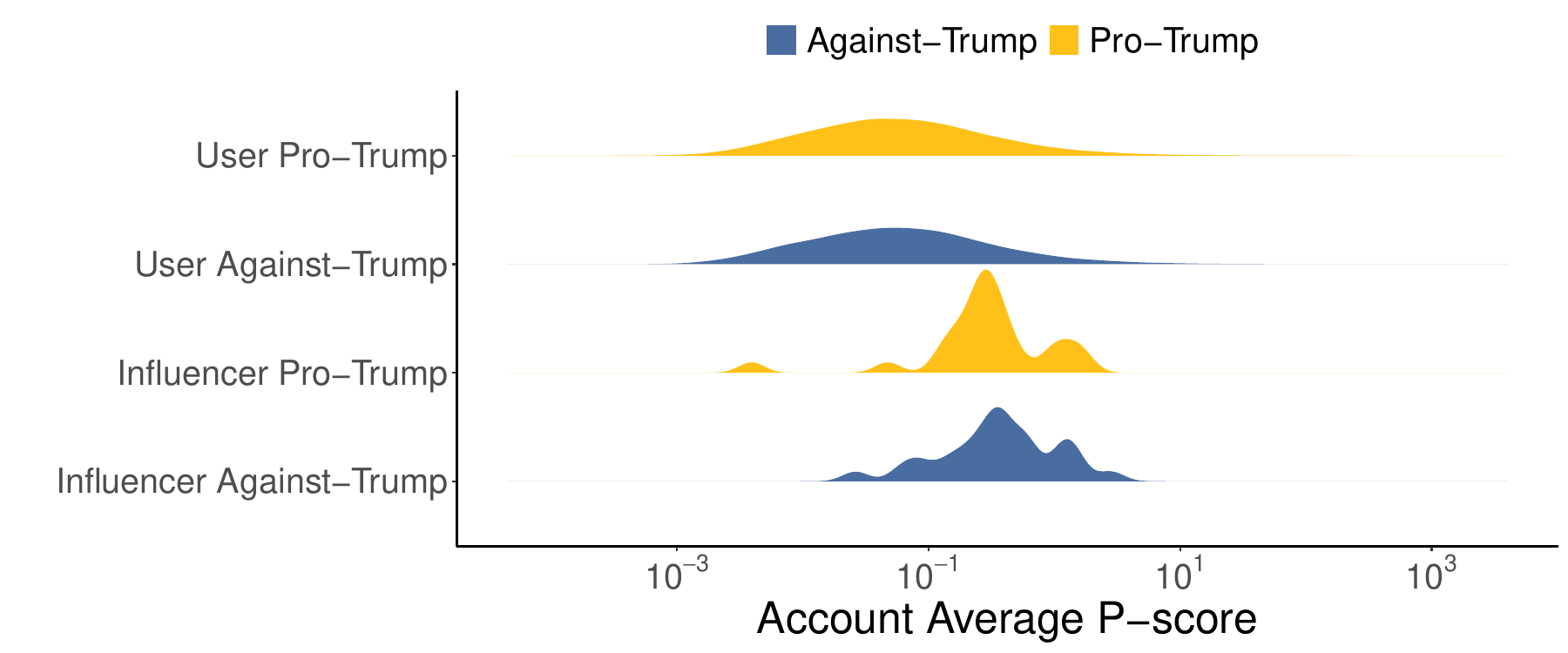}
    \caption{2024 US Elections}
    \end{subfigure}
    \caption{Average p-score distributions for users and influencers opposing and supporting military aid to Ukraine and Donald Trump.}
    \label{fig:avg_pscore_distribution}
\end{figure*}

This is indicative of similar visibility levels across different thematic content and suggests the lack of interventions at the topic level.
We also investigate the distribution of users within each interval by plotting the Gini index on the distribution of the number of tweets per interval for each author.
Figure~\ref{fig:disparity_lambretta} reports the distribution of p-scores across tweet categories, with red lines indicating the median value for each category, alongside the Gini index for each interval for the Ukraine-Russia and the 2024 US Elections datasets.
For the Ukraine-Russia dataset, the median p-score for the different topical themes is very close to each other (within a Standard Deviation of 0.0296), with the topic ``War Related Reporting and Events'' having the highest median (0.143), and ``Criticism of US Military Support to Ukraine'' the lowest median p-score (0.071828).
A similar pattern is observed in the US Elections dataset, where the highest median p-score corresponds to the ``International Conflicts \& Electoral Impact'' theme (0.0561), while the lowest to the ``Policy Debates \& Issues'' theme (0.0349).
Taken together, these results suggest the lack of any topic-oriented intervention on the algorithm's part.

\subsection{RQ2: User Level}
Besides content-level visibility, reductions can also affect individual accounts, regardless of what they post.
Hence, we analyze whether users experience different levels of visibility based on their role in the debate and their ideological stance, specifically, their support for or opposition to military aid to Ukraine as well as support for or opposition to Donald Trump in the lead-up to the 2024 US Elections.

\descr{Inferring Users' Ideology.} To perform account-level analysis, we first need to infer users' attributes, such as their stance on the debated issue.
To this end, we use latent ideology estimation (see Section~\ref{sec:latent}) to classify users' ideological stances.
Given a set of prominent accounts known as influencers (see Section~\ref{sec:influencers}), this technique allows us to infer users' stances based on interaction information.
Matrix decomposition maps users who interacted with a similar set of influencers close to each other in a real space, while users with fewer common influencers are positioned farther apart.

Due to the absence of retweet statistics for the 2024 US Elections dataset, we estimate latent ideology using retweet interactions for the Ukraine-Russia and reply interactions for the US Elections datasets.
\revision{C6}{As mentioned, we do so as the latter only includes a few hundred retweets that did not have a valid view count.}
\revision{C6}{Although users' estimaton results a bit noisier, the use of replies does not affect the overall ability to infer accounts' stances.}
Indeed, the bar plots in Figure~\ref{fig:ideology_analysis} show the distribution of ideology scores for users and influencers in the Ukraine–Russia War (panel a) and the US Elections (panel b), together with the distribution of ideology scores for repliers/retweeters of a subset of ten influencers in each debate.
Both datasets exhibit two communities positioned at opposite ends of the $[-1,1]$ space, revealing polarized debates with bimodal opinion distributions, as also highlighted by Baqir et al.~\cite{baqir2025unveiling} in the context of the Ukraine-Russia war \revision{}{and by the Dip test as reported in Section~\ref{sec:influencers}}.

In each debate, we categorize accounts into two stances: supporters (ideology score $<0$) or opponents of military aid (ideology score $>0$), and supporters (ideology score $>0$) or opponents of Donald Trump (ideology score $<0$).
We then calculate the average p-score for users and influencers in our dataset.
To ensure the robustness of our results, we only consider accounts with more than five original tweets.

\descr{Visibility vs Ideology.}
After inferring user opinions, we analyze whether the visibility of the account is influenced by the ideology they support.
Figure~\ref{fig:avg_pscore_distribution} illustrates the distributions of the average p-score for users and influencers representing opposing ideological stances in the two debates.
Influencers tend to have higher average p-scores than users in both debates, likely due to their prominent role and greater popularity.
However, there is little difference between the p-score distributions of users and influencers when comparing ideological stances in both the Ukraine-Russia and US Elections datasets.
For the Ukraine-Russia dataset, the median p-scores are 0.118 for supporters vs.~0.136 for opponents among users, and 0.479 for supporters vs.~0.499 for opponents among influencers.
For the US Elections dataset, the median p-scores are 0.049 for Trump supporters vs.~0.048 for opponents among users, and 0.280 for Trump supporters vs.~0.358 for opponents among influencers.
These results are confirmed by the Mann-Whitney U test, which shows significantly higher performance for influencers compared to users (p-value $<$ 0.0005 for both supporter and opponent influencers versus users; p-value $< 1.30e-07$ for pro- and anti-Trump influencers versus users).
The test also reveals comparable visibility between users and influencers of different stances (p-value $>$ 0.07 for the Ukraine-Russia case and p-value $>$ 0.29 for the US elections).

\begin{figure*}[t]
    \centering
    \begin{subfigure}[t]{0.95\linewidth}\vskip 0pt
    \includegraphics[width=\linewidth]{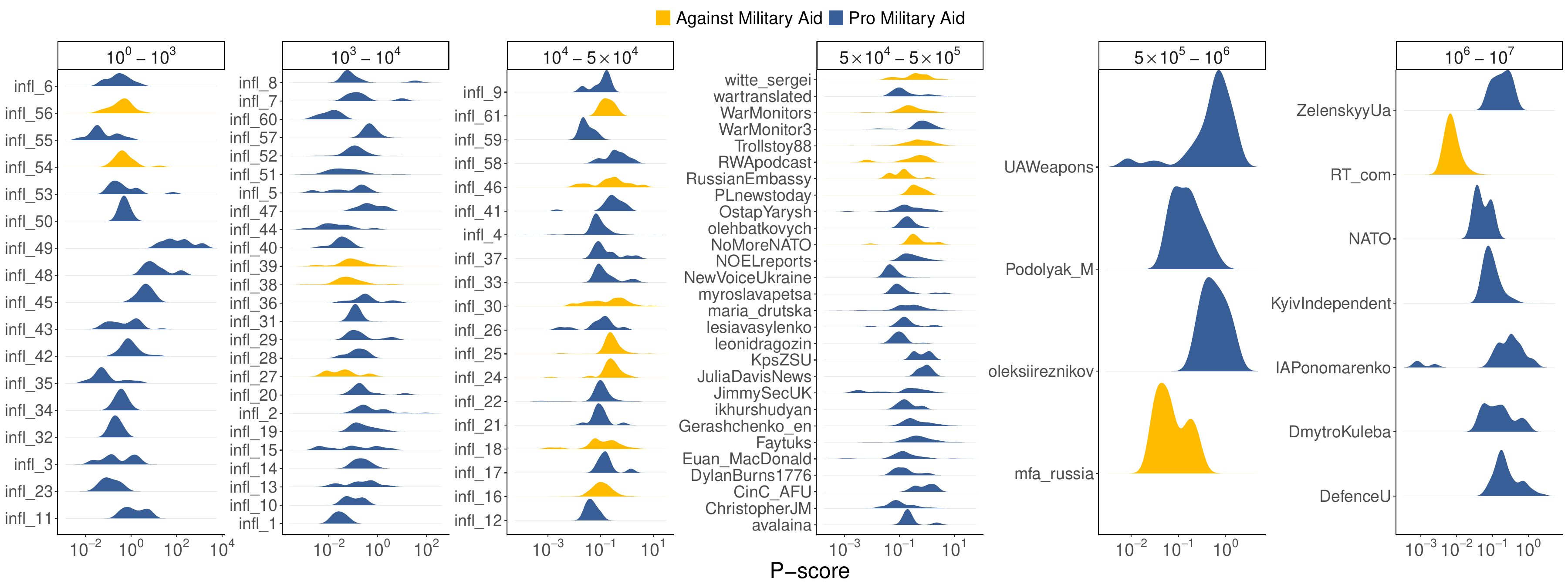}
    \caption{Ukraine-Russia War}
    \end{subfigure}\\
    \begin{subfigure}[t]{0.999\linewidth}\vskip 0pt
    \includegraphics[width=\linewidth]{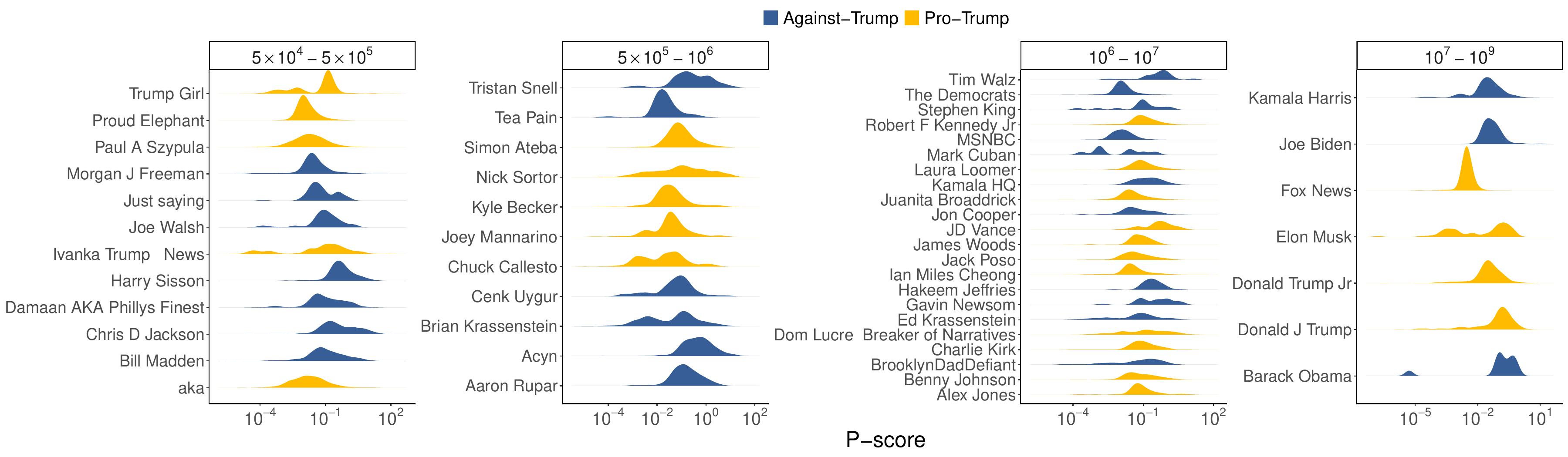}
    \caption{2024 US Elections}
    \end{subfigure}
    \caption{Distribution of p-score for influencers according to their popularity (i.e., number of followers). Only influencers with more than five original tweets were considered, and accounts with fewer than 50,000 followers were anonymized.}
    \label{fig:influencers_distributions}
\end{figure*}

\descr{Influencers p-score distributions.}
Although we have ruled out the penalization of specific ideologies, visibility alterations may still occur for specific accounts.
This is suggested by the presence of outliers among influencers in Figure~\ref{fig:avg_pscore_distribution}, which may indicate accounts experiencing either reduced or increased content circulation due to algorithmic visibility adjustments.
To investigate the presence of algorithmic interventions and to mitigate the influence of potential confounding factors, we exclusively focus on influencers and divide them based on their popularity in different tiers, as done in recent work~\cite{tricomi2024climbing}.
Figure~\ref{fig:influencers_distributions} displays the p-score distribution for the accounts used as influencers in the analysis of the two debates that posted more than five original pieces of content, with accounts grouped according to the popularity tier they belong to.

For both datasets, within the same group, influencers tend to exhibit distributions with similar shapes but different medians, reinforcing the possibility of algorithmic intervention at the individual level.
We also observe that some influencers exhibit multiple peaks in their p-score distributions.
This pattern could be attributed to the use of different types of content, each achieving varying levels of visibility, as illustrated in Figure~\ref{fig:p_score_all_content}.
Alternatively, it may reflect temporal fluctuations in visibility due to adjustments in the suggestion algorithm.
Notably, in the Ukraine-Russia war debate, mfa\_russia and RT\_com (accounts associated with the Ministry of Foreign Affairs and a Russian state-sponsored news outlet, respectively) significantly underperform relative to other accounts within the same tier.
In the US Elections debate, the accounts linked to Joe Biden and Kamala Harris exhibit p-score distributions skewed lower than that of Donald Trump.

\descr{Case Studies.}
To shed light on the individual differences observed earlier, we select pairs of users with comparable roles in the two debates and analyze the distribution of their p-scores.
In the case of the Ukraine-Russia war debate, we analyze RT\_com and compare it to KyivIndependent, a news outlet supporting Ukraine.
We select these accounts as they have comparable levels of popularity and belong to news agencies representing opposing factions, thereby minimizing potential confounding factors that could influence visibility.
For the 2024 US Elections, we select the final candidates from the two major parties, i.e., Donald Trump and Kamala Harris.

\begin{figure}[t]
    \centering
    \begin{subfigure}[t]{0.485\columnwidth}\vskip 0pt
    \includegraphics[width=\linewidth]{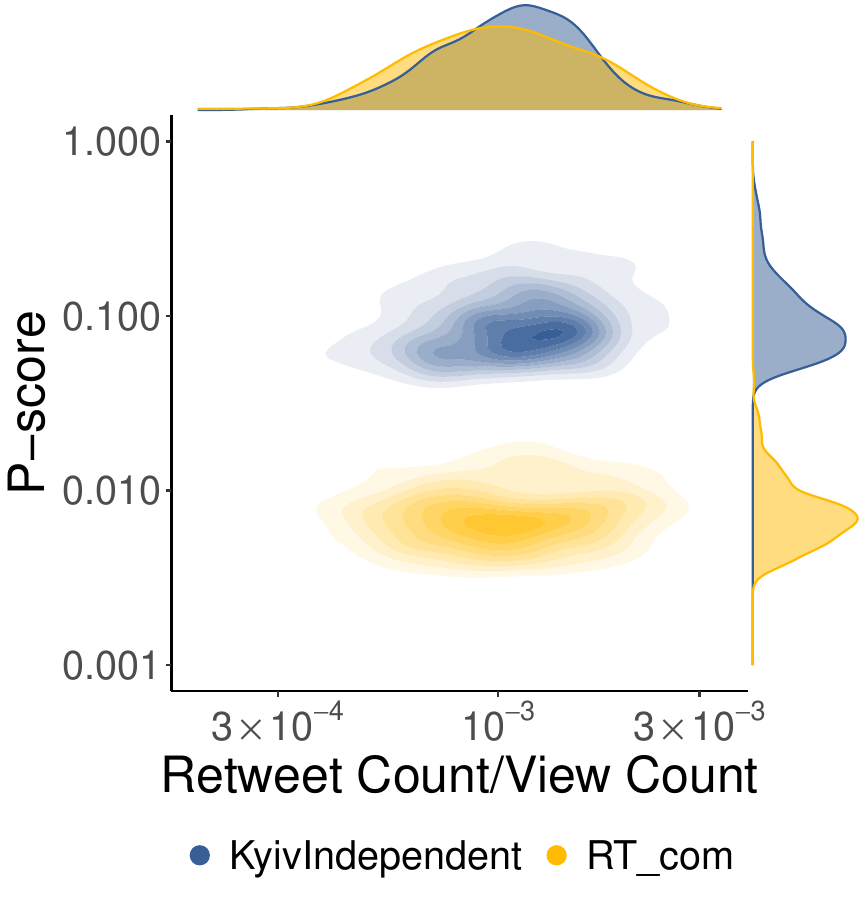}
    \caption{Ukraine-Russia War}\label{case-ukraine}
    \end{subfigure}
    \hspace{-0.25cm}
    \begin{subfigure}[t]{0.5\columnwidth}\vskip 0pt
    \includegraphics[width=\linewidth]{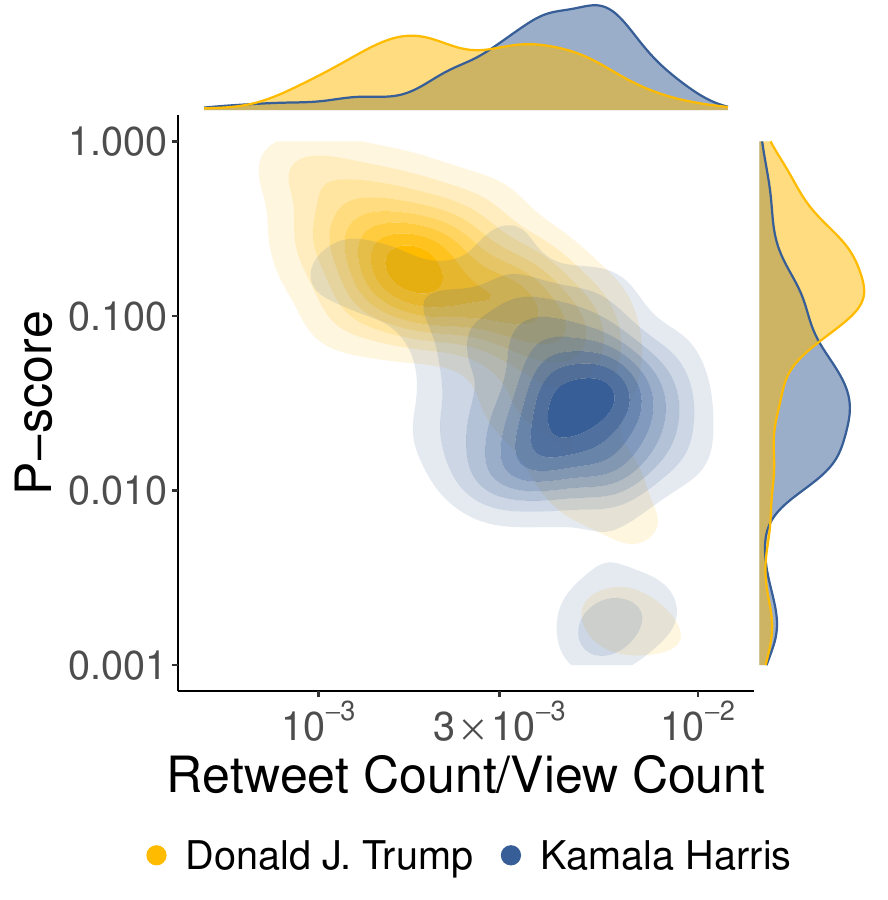}
    \caption{2024 US Elections}\label{case-elections}
    \end{subfigure}
    \caption{P-score vs.~retweets per impression count for content by two news agencies with opposing ideological stances. %
    The volume of content produced by the four accounts is 553 (KyivIndependent), 258 (RT\_com), 223 (Donald J. Trump), and 467 (Kamala Harris).} %
    \label{fig:kyiv_rt}
\end{figure}

One might argue that the circulation of content also depends on how much other users engage with it; thus, a more active user base could enhance the diffusion of the author's content.
To assess this, Figure~\ref{fig:kyiv_rt} presents a 2-D density plot of the p-score versus the ratio of retweets to views.
This ratio serves as a proxy for quantifying how actively consumers re-share content after being exposed to it.
In particular, Figure~\ref{case-ukraine} %
reveals that RT\_com and KyivIndependent have rather similar user bases in terms of active sharing, as highlighted by the top marginal distribution (Mann-Whitney U test p-value: 0.093).
However, they display significant differences in their p-score distributions (Mann-Whitney U test p-value: $< 2.2e-16$,  median p-score: KyivIndependent=0.084, RT\_com=0.007).
Given the similarities across various dimensions, these findings strongly support the hypothesis that content from RT\_com has been subject to visibility reduction.

\begin{figure*}[t!]
    \centering
    \begin{subfigure}[t]{0.49\linewidth}\vskip 0pt
    \includegraphics[width=\linewidth]{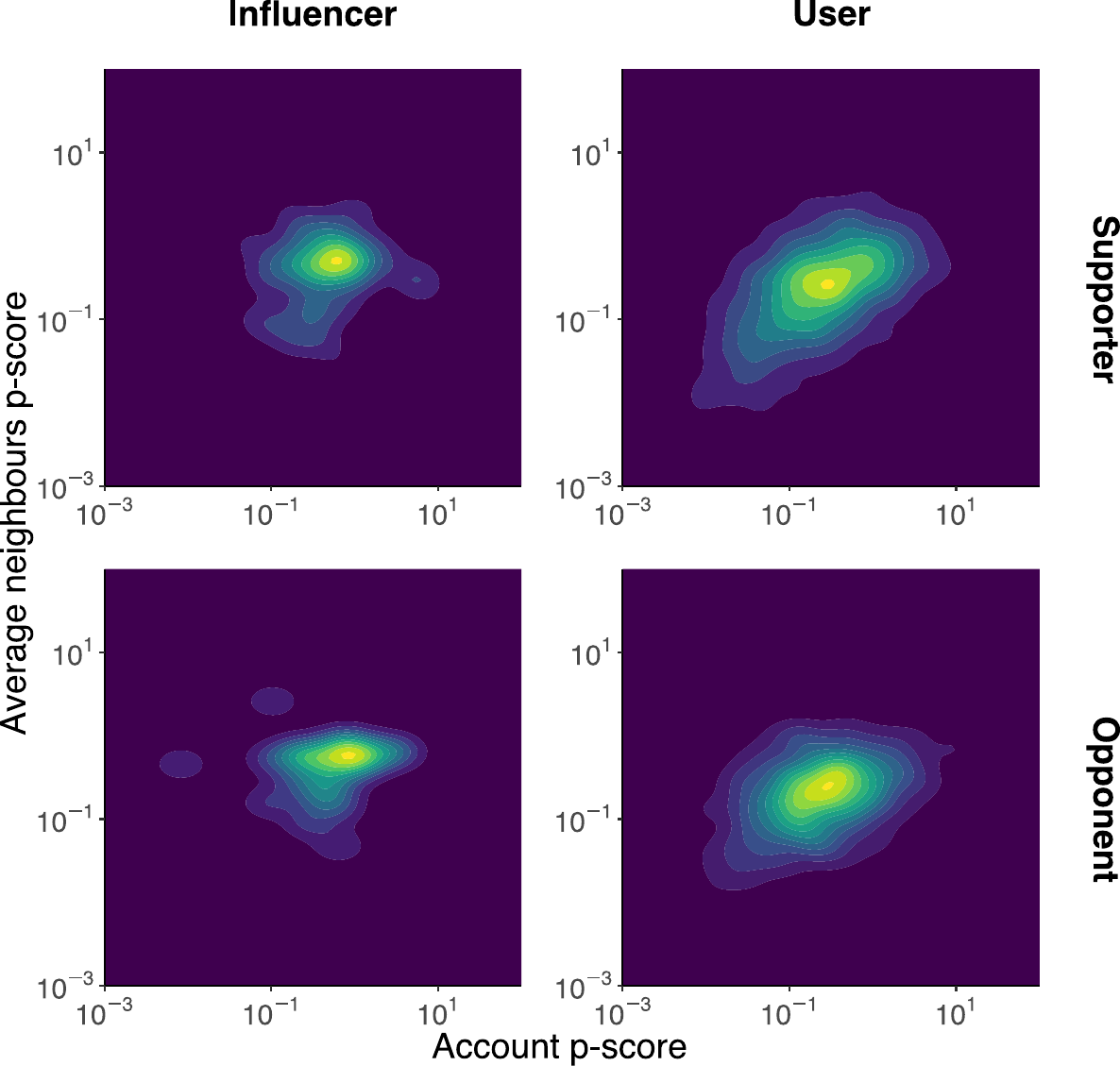}
    \caption{Ukraine-Russia War}\label{density-ukraine}
    \end{subfigure}
    ~
    \begin{subfigure}[t]{0.49\linewidth}\vskip 0pt
    \includegraphics[width=\linewidth]{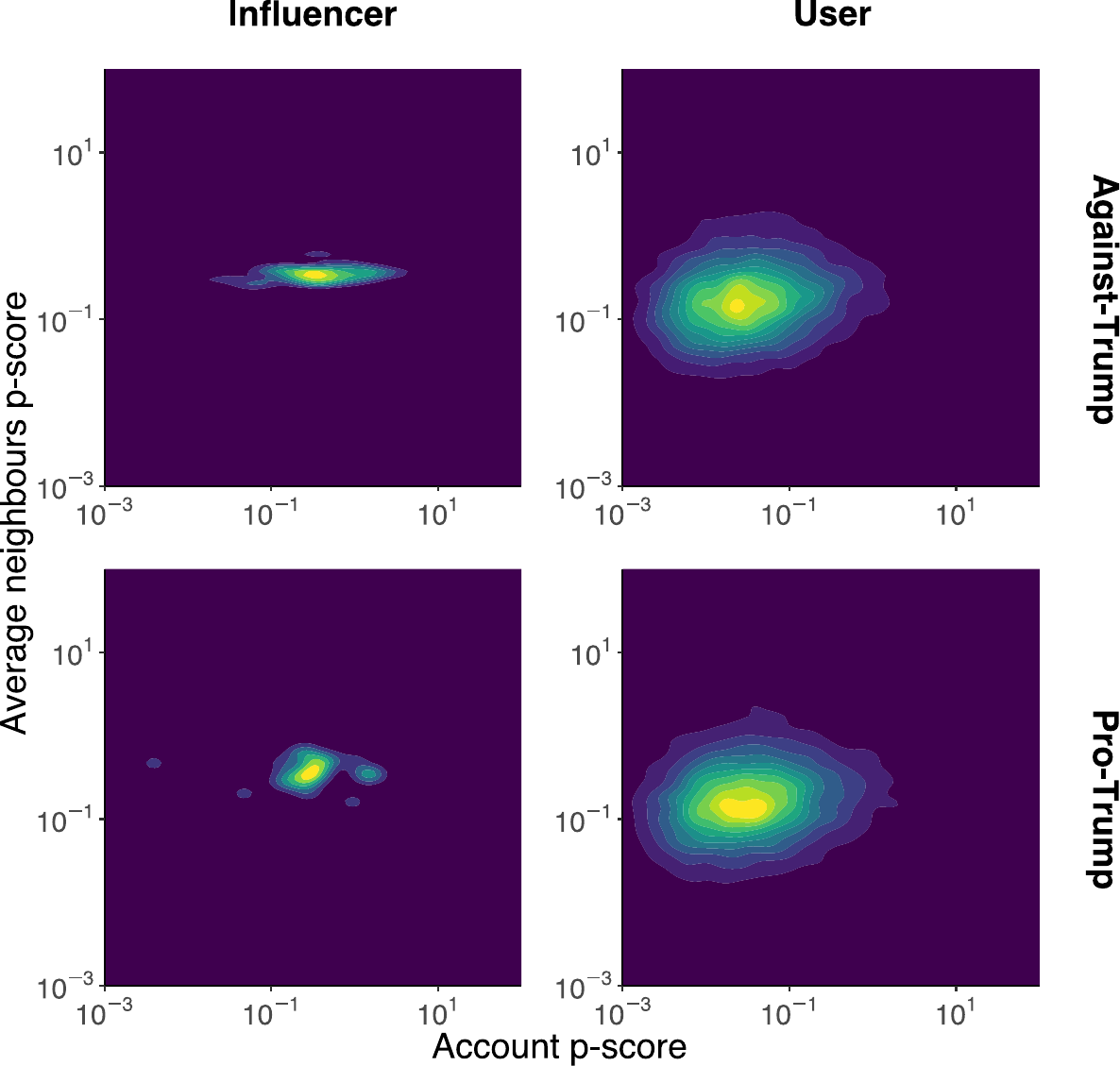}
    \caption{2024 US Elections}\label{density-elections}
    \end{subfigure}
    \caption{Joint density distribution of account p-scores versus the average p-score of the account that interacted with them, categorized by different stances and types of accounts. For the Ukraine-Russia war, the interaction network was constructed using retweets, while, for the US Elections, we used replies.}
    \label{fig:nn_score}
\end{figure*}

In the case of the US Elections, the separation between Trump-Harris distributions is less clear (Figure~\ref{case-elections}), as the two 2D densities overlap in some areas.
Nevertheless, their marginal distributions differ significantly along both dimensions.
Specifically, the distribution of retweet count over view count reveals that Donald Trump's audience is less prone to explicitly endorse his content through active engagement compared to Kamala Harris's audience (median retweets/views: 0.0025 for Trump, 0.0043 for Harris, Trump vs Harris Mann-Whitney U test p-value=1, Harris vs Trump Mann-Whitney U test p-value $= 4.543e-15$).
However, the p-score marginal distributions show that most of the content produced by Donald Trump achieves greater visibility than that produced by Kamala Harris (p-score median: 0.13 Trump, 0.03 Harris, Trump vs Harris Mann-Whitney U test p-value $< 2.2e-16$).
Hence, despite having a less active user base, Donald Trump's content experienced increased visibility, suggesting the possible influence of algorithmic intervention.

\subsection{RQ3: Network Level}
Last but not least, to answer RQ3, we build retweet and reply interaction networks and study whether users interacting with each other experience similar visibility.
We compute the joint density of the p-score for each account with respect to the average p-score of the accounts that interacted with it, as illustrated in Figure~\ref{fig:nn_score}.
To avoid potential conflating factors related to different roles, %
we compute separate density distributions for influencers and users based on their ideological stance.
For the Ukraine-Russia war (Figure~\ref{density-ukraine}), we use retweet information to capture interaction, while, for US Elections (Figure~\ref{density-elections}), we employ replies to build the interaction network due to the unavailability of retweets.

The analysis indicates an absence of correlation between individual p-scores and the average p-scores of retweeters/\revision{}{repliers} across both account classes and ideological stances for both datasets.
Thus, users who interacted with accounts with reduced visibility do not necessarily experience decreased visibility themselves, regardless of account type, stance, or debate.
This is further supported by Pearson's correlation test, which yielded a maximum correlation value of 0.13 for the Influencer-Supporter case in the Ukraine-Russia dataset and a value of 0.094 in the US elections debate.

\section{Discussion}
\label{sec:discussion}
The visibility of content online is inherently shaped by recommendation algorithms, as these determine what users see based on various engagement metrics and platform-specific goals~\cite{jiang2020reasoning}. 
Many platforms report reducing visibility as part of their moderation strategies to limit the spread of harmful or borderline content, purportedly to safeguard public discourse~\cite{gillespie2022not}. 
\revision{C2c}{However, algorithmic visibility reduction may occur for several reasons, for instance, to penalize low-performing content or to prioritize user retention.}

\revision{}{Regardless,} the lack of transparency around these interventions contributes to a growing distrust towards platforms~\cite{savolainen2022shadow}.
Moreover, while interventions can reduce the spread of harmful content or limit the influence of problematic users (including bots or coordinated disinformation campaigns), there is no guarantee that the same mechanisms are not used to silence minority voices or suppress narratives contrary to corporate interests. 
These interventions are subtle, far less noticeable than bans or content removal, and users may remain unaware that the reach of their posts is being reduced~\cite{gillespie2022not}.

Our work investigated visibility reduction in the context of two Twitter datasets related to two polarized topics (the Ukraine-Russia war and the 2024 US Elections) using the content views metric. 
This allowed us to directly observe the effects of algorithmic interventions and assess how visibility reductions vary based on several factors. 
In doing so, we analyze both the behavior of recommendation algorithms at a granular level and how views can help identify otherwise unnoticed visibility suppression.

\descr{Data availability.} Practices like shadow banning can significantly influence online debates and shape public opinion.
As a result, the ability to study these interventions is crucial. 
However, while data availability is essential in doing so, in recent years, access to social media data for research purposes has become increasingly restricted~\cite{kupferschmidt2023twitter}, exacerbating the opacity of platforms and further eroding public trust.
For instance, there is growing concern that social media companies could reduce content visibility to reduce the spread of misinformation, but may be choosing not to do so~\cite{bagchi2024social}.

Indeed, our study reveals that Twitter did not systematically apply visibility reduction techniques to limit the spread of unreliable sources both in the Ukraine-Russia war and the 2024 US Presidential Elections debates. 
Instead, interventions appear to target specific accounts flagged as problematic, while all content containing external links has been penalized, regardless of the credibility of the information shared.

\descr{Longitudinal analysis.} The comparison between multiple datasets collected during different periods and on different topics also helped us highlight the evolution of visibility depending on the context considered.
For example, although the algorithm tends to reduce the visibility of tweets containing URLs, the strength of such interventions appears to differ between the two cases.
Similarly, the differences in the visibility analysis at the individual level may result from variations in the behavior of the recommendation algorithm.

\descr{Additional factors.} Finally, other factors influencing content visibility, including coordinated behaviors, user engagement strategies, and the topic under discussion, should also be taken into consideration. 
While these factors can exploit algorithmic design to boost visibility, our methodology proves effective in detecting visibility disparities between types of content and user behavior independently from the origin of these differences. 
\revision{C1}{In other words, alternative explanations for our findings are, arguably, far less likely.

For instance, one possible reason for visibility alteration could be related to bots, however,  our visibility metric adjusts for follower count to account for potential bot-driven inflation. 
Furthermore, our network analysis is designed to detect clusters of users with altered visibility, including those that may be the result of coordinated behavior. 
Since no penalized communities were identified, such groups, if they exist, either evade detection or do not benefit from increased visibility, making coordinated manipulation an unlikely explanation. 
At the user level, we include retweets to capture engagement-driven visibility; nevertheless, users with similar roles and activity levels still exhibit significant differences in visibility, which suggests that user-level factors alone cannot explain the observed penalization. 
Lastly, the systematic reduction in visibility of content containing URLs in both debates, regardless of the type of resource referenced, strongly suggests algorithmic intervention rather than varying user interest.}

\descr{Ethics Considerations.}
Our work relies on publicly available data obtained from other studies~\cite{baqir2025unveiling,balasubramanian2024public}, retrieved using the Twitter public API or by scraping public data, and augmented with publicly available tools. 
As such, it is not considered human subject research by our institutions. 
\revision{C2b}{We acknowledge that one of our datasets was collected by~\cite{balasubramanian2024public} using a custom scraper, which may potentially contravene Twitter's terms of service (ToS).
However, we believe our research to be in the public interest, as studying platform manipulation arguably outweighs ToS concerns, in a manner not dissimilar to, e.g., violating robots.txt when crawling cybercrime forums~\cite{brewer2021ethics}.
In general, ethics researchers have made the case that ethics might not depend on breaking the ToS of the platforms~\cite{fiesler2020no}.
}

Furthermore, to preserve anonymity, we do not report any information about the users in the dataset other than those of public entities like news organizations. 
For accounts that play an important role in the debate (i.e., influencers), we anonymize those with fewer than 50,000 followers, aligning with previous research~\cite{falkenberg2022growing}.
\revision{C2b}{Overall, we believe that our work aligns with the Menlo Report~\cite{menlo}'s principles of Beneficence, Respect for Persons, and Public Interest.
}

\section{Conclusion}
This paper examined differences in content visibility on Twitter/X -- at content, user, and network levels -- in the context of the Ukraine-Russia war and the 2024 US Presidential Elections.
\revision{C4}{We relied on view counts, a largely unexplored feature of Twitter/X posts, to estimate content circulation and detect forms of visibility penalization that are not captured by techniques previously used in the literature.}
We found evidence that Twitter/X consistently favors tweets that do not contain links to external websites. 
While the platform does not systematically alter content visibility based on political views, the factuality of the referenced sources, or ideological stances, we do find evidence of visibility alterations at the individual account level, as illustrated by the cases of RT.com versus The Kyiv Independent and Donald Trump versus Kamala Harris.
Our work demonstrates how visibility alterations can be detected using engagement metrics like view counts, an underexplored and often unavailable metric.

\revision{C2d}{As mentioned, our work mainly focuses on algorithmic penalization in a broad sense, while we do not aim to determine whether/how this may be ideologically motivated.}

\descr{\revision{C2f}{Recommendations.}}
\revision{}{We believe that the methods proposed in this paper can help researchers and civil society alike improve transparency by uncovering penalization practices adopted by social media platforms, especially if they might end up infringing on free speech.
Concretely, non-profits, research outfits, and fact-checkers can leverage our methods to raise awareness and increase transparency by publishing dashboards of content, keywords, and domains that are anomalously penalized, and record historical information about potential changes in visibility dynamics in public reports.}

\descr{Limitations.} Our study revolves around two datasets covering two different topics and collected using different methods, i.e., the Academic API and Web scraping.
This may result in varying levels of representativeness of the debate. 
To mitigate this potential issue, we pre-processed the data to ensure strong coherence and minimize the likelihood of any artifacts influencing our results.
Also, although our methodology proved effective in detecting visibility alterations, future and/or more sophisticated strategies may still go undetected.

Finally, going forward, concerns about the limited availability of social network data for researchers cast doubts on the feasibility of conducting at-scale audits and studies of algorithmic behavior and practices adopted by platforms.
Without data, ensuring accountability and maintaining trust in these platforms will remain an uphill battle.
Initiatives like the Digital Services Act (DSA)~\cite{dsa} aim to enforce greater transparency, however, the real-world implementation of these laws may prove challenging.

\descr{Future Work.} Our work could be enhanced in several ways. 
Integrating other factors -- e.g., the influence of bots and coordinated accounts -- could provide a more thorough understanding of how these elements affect visibility and how the algorithm responds to their presence. 
Also, as part of our data was retrieved using official APIs, we acknowledge that platforms may influence the data they provide to mask intervention patterns. 
While we took several integrity-enhancing actions to ensure the reliability of the data, one could combine API-based and scraping techniques to collect the same dataset to offer even more robust and nuanced insights.

Moreover, a comparison across additional datasets covering additional topics over extended periods could help identify changes in algorithmic behavior.
Finally, comparing visibility patterns across other platforms would tease out unique traits of Twitter's recommendation algorithm while also studying practices shared with other systems.

\descr{Acknowledgments.}
A.G. gratefully acknowledges the financial support provided by CY4GATE, by the National Recovery and Resilience Plan (NRRP) project “Securing Software Platforms – SOP” (CUP H73C22000890001), and by the European Union under the NRRP, Mission 4 Component 2 Investment 1.3 – Call for Proposals No. 341 of March 15, 2022, Italian Ministry of University and Research – NextGenerationEU, Project Code PE00000014, Concession Decree No. 1556 of October 11, 2022 (CUP D43C22003050001), “SEcurity and RIghts in the CyberSpace (SERICS) – Spoke 2 Misinformation and Fakes: DEcision supporT systEm foR cybeR intelligENCE (Deterrence).
G.S.'s work was partially supported by NSF Grant CNS-2419829. 

\small
\bibliographystyle{abbrv}

\end{document}